\documentclass[onecolumn,twocolumn, journal]{IEEEtran}
\usepackage[T1]{fontenc}
\usepackage[latin9]{inputenc}
\usepackage{color}
\usepackage{bm}
\usepackage{amsmath}
\usepackage{amsthm}
\usepackage{amssymb}
\usepackage{graphicx}
\usepackage[bookmarks=true,bookmarksnumbered=true,bookmarksopen=true,bookmarksopenlevel=1,
 breaklinks=false,pdfborder={0 0 1},backref=false,colorlinks=false]
 {hyperref}
\hypersetup{pdftitle={Your Title},
 pdfauthor={Your Name},
 pdfborderstyle=,pdfpagelayout=OneColumn,pdfnewwindow=true,pdfstartview=XYZ,plainpages=false}

\makeatletter

\providecommand{\tabularnewline}{\\}

\theoremstyle{plain}
\newtheorem{thm}{\protect\theoremname}

\providecommand{\theoremname}{Theorem}
\usepackage{amsthm}
\newtheorem{assumption}{Assumption}

\newtheorem{definition}{Definition}
\newtheorem{lemma}{Lemma}
\usepackage{cite}
\usepackage{hyperref}
\hypersetup{
  colorlinks=true,   
  linkcolor=blue,    
  citecolor=blue,   
  urlcolor=blue      
}

\ifdefined\showcaptionsetup
 \PassOptionsToPackage{caption=false}{subfig}
\fi
\usepackage{subfig}
\makeatother

\providecommand{\theoremname}{Theorem}

\begin{document}
\title{Unsupervised Deep Equilibrium Model Learning for Large-Scale Channel
Estimation with Performance Guarantees}
\author{Haotian~Tian,~\IEEEmembership{Student~Member,~IEEE,} and~Lixiang~Lian,~\IEEEmembership{Member,~IEEE}\thanks{The material in this paper was presented in part at the IEEE Global
Communications Conference (GLOBECOM), Cape Town, South Africa, Dec.
2024 \cite{tian2024gsure}. \emph{(Corresponding author: Lixiang Lian.)}}\thanks{Haotian~Tian and Lixiang Lian are with the School of Information
Science and Technology, ShanghaiTech University, Shanghai 201210,
China (e-mail: \{tianht2022, lianlx\}@shanghaitech.edu.cn).}}
\maketitle
\begin{abstract}
\label{abstract}Supervised deep learning methods have shown promise
for large-scale channel estimation (LCE), but their reliance on ground-truth
channel labels greatly limits their practicality in real-world systems.
In this paper, we propose an unsupervised learning framework for LCE
that does not require ground-truth channels. The proposed approach
leverages Generalized Stein's Unbiased Risk Estimate (GSURE) as a
principled unsupervised loss function, which provides an unbiased
estimate of the projected mean-squared error (PMSE) from compressed
noisy measurements. To ensure a guaranteed performance, we integrate
a deep equilibrium (DEQ) model, which implicitly represents an infinite-depth
network by directly learning the fixed point of a parameterized iterative
process. We theoretically prove that, under mild conditions, the proposed
GSURE-based unsupervised DEQ learning can achieve oracle-level supervised
performance. In particular, we show that the DEQ architecture inherently
enforces a compressible solution. We then demonstrate that DEQ-induced
compressibility ensures that optimizing the projected error via GSURE
suffices to guarantee a good MSE performance, enabling a rigorous
performance guarantee. Extensive simulations validate the theoretical
findings and demonstrate that the proposed framework significantly
outperforms various baselines when ground-truth channel is unavailable.
\end{abstract}


\IEEEpeerreviewmaketitle{}

\section{Introduction}

\IEEEPARstart{N}{ext-generation} communication systems like 6G and
beyond will feature large-scale antenna arrays, large reconfigurable
intelligent surfaces (RISs), ultra-dense networks and high-density
users, enabling advanced applications such as terahertz (THz) communication,
holographic communication, RIS-assisted communication, the massive
Internet of Things (IoT), etc. The large-scale nature of the underlying
channels underscores the importance of accurate and robust large-scale
channel estimation (LCE) in achieving the desired performance and
reliability in the emerging technologies.

Deep learning (DL) has emerged as an effective tool for the LCE task
\cite{hu2020deep,dong2019deep,borgerding2017amp,wei2020deep,ma2021model,yu2023adaptive,zheng2021online,lian2023regularized,zhang2023self,he2022beamspace,yu2024bayes},
whose performance has surpassed many traditional optimization-based
algorithms, establishing itself as the mainstream research direction.
Although supervised channel learning (SCL) can achieve accurate channel
estimates \cite{hu2020deep,dong2019deep,borgerding2017amp,wei2020deep,ma2021model,yu2023adaptive},
it relies on large volumes of ground-truth channel data, which are
usually hard to measure and hardware-intensive to acquire due to their
dynamic nature and large scale. This motivates the study of unsupervised
channel learning (UCL). Since UCL eliminates the reliance on ground-truth
channels, it can easily adapt to dynamically changing environments
by updating the model based on real-time received measurements. However,
the lack of labeled data during training may lead to a self-directed
learning process, resulting in unreliable channel estimates without
performance guarantees. 

\subsection{Related Work}

\subsubsection{Supervised LCE}

Early works in SCL primarily focused on learning the mapping between
measurements and channel data by minimizing the mean-squared error
(MSE) loss function. For example, a fully-connected ReLU deep neural
network (DNN) was deployed in \cite{hu2020deep} to estimate channels
for single-input multiple-output (SIMO) systems. \cite{hu2020deep}
demonstrates that the DNN approximates the minimum mean-squared error
(MMSE) estimator when both the number of hidden layers and the dataset
size are sufficiently large. Similarly, a deep convolutional neural
network (CNN) was adopted in \cite{dong2019deep} for channel estimation
in massive multiple-input-multiple-output (MIMO) systems. To enhance
the learning efficiency and improve the interpretability of SCL, some
specialized networks were employed by incorporating ideas from traditional
iterative LCE algorithms.

Deep unrolling (DU) has been widely adopted to mimic the "basic"
iterative compressed sensing (CS) algorithms for LCE with a limited
number of iterations \cite{borgerding2017amp,wei2020deep,ma2021model}.
However, to avoid the numerical instability in the training process
and alleviate the storage costs, the number of unrolled iterations
is typically difficult to sufficiently increase, which significantly
limits the performance of DU. To circumvent this challenge, an alternative
architecture, deep equilibrium model (DEQ), has been adopted in \cite{yu2023adaptive}
for terahertz channel estimation. Unlike DU, DEQ, originally proposed
in \cite{bai2019deep}, directly learns the fixed point of a nonlinear
transform $\mathbf{x}^{\star}=f_{\mathbf{\Theta}}(\mathbf{x}^{\star};\mathbf{y})$
by constructing a fixed-point iteration and incorporating the update
structure of a classical optimization algorithm into the network model.
Moreover, compared to DU, which needs to store the results of each
layer to perform backpropagation, the backpropagation in DEQ can be
conducted at fixed point through implicit differentiation, and thus
only constant memory is required \cite{fung2022jfb,gilton2021deep}.
Leveraging the theoretical insights from traditional iterative algorithms,
the SCL based on DU and DEQ allows for convergence analysis \cite{chen2018theoretical}
by establishing a correspondence with the underlying iterative algorithms.

\subsubsection{Unsupervised LCE}

To eliminate the reliance on the ground-truth channel data, there
have been several attempts to design UCL. The authors in \cite{zheng2021online}
proposed to train a DNN by minimizing online loss functions such as
least square (LS) and nuclear norm-based loss functions. Convergence
analysis was provided based on the universal approximation theorem
(UAT) of DNN \cite{Calin2020}. Although UAT guarantees that fully
connected structures can approximate complex functions, it requires
large-scale models and substantial amounts of training labels to meet
performance requirements. Moreover, the theorem is derived under ideal
conditions, offering little practical guidance on network architecture
design or parameter learning, and the resulting models often lack
interpretability. Recent studies have examined UCL based on Stein's
unbiased risk estimate (SURE) in \cite{he2022beamspace} and Stein's
score function in \cite{yu2024bayes}. However, these methods are
developed under the assumption of a white additive Gaussian noise
(AWGN) at each step of their algorithms, which leads to two inherent
limitations. First, the AWGN assumption in each step may not always
hold when estimation conditions are unmet. Second, the equivalent
AWGN variance varies with the iteration index and lacks a reliable
method for precise estimation. Deep image prior has been employed
in \cite{lian2023regularized} to generate large-scale channels in
real-time by leveraging the sparse structure of the channels in an
unsupervised manner. A CNN with residual connections is utilized in
\cite{zhang2023self} for self-supervised channel learning in RIS-aided
communication systems, without the need for ground truth. However,
these works generally lack theoretical guarantees regarding the performance
of UCL methods.

\subsection{Our Contributions}

Some SCL algorithms have demonstrated near-MMSE optimality due to
their access to extensive ground truth data \cite{metzler2017learned,borgerding2017amp,chen2018theoretical,hu2020deep}.
In contrast, UCL faces significant challenges in offering guaranteed
performance due to the absence of labels, difficulty in model evaluation,
high risk of poor convergence, etc. To ensure performance guarantees
in UCL, two key elements are required: (i) network architectures that
promote learning of meaningful structure, and (ii) well-designed proxy
losses that guide training in the absence of labels. In this paper,
we address LCE from noisy compressed measurements via an unsupervised
DEQ-based framework. DEQ is adopted for its ability to directly learn
the fixed point of a nonlinear mapping with constant memory. To ensure
convergence and promote sparse structure learning, we enforce a contraction
property for DEQ. Furthermore, we employ Generalized Stein's Unbiased
Risk Estimate (GSURE) \cite{eldar2008generalized} as the unsupervised
loss function. Through theoretical analysis, we show that the proposed
GSURE-based unsupervised DEQ learning can approach oracle-level performance,
i.e., the performance limit of DEQ attainable by supervised learning
with ground truth data. The main contributions are summarized as
follows:
\begin{itemize}
\item \textbf{GSURE-based Unsupervised DEQ Learning (GUDL):} We formulate
the LCE as a sparse signal recovery problem and propose GUDL to solve
it. In particular, DEQ is employed to learn the fixed point of a proximal
algorithm and GSURE serves as an unsupervised loss function for DEQ
training. We provide convergence guarantees for the DEQ forward pass.
\item \textbf{Theoretical Performance Analysis: }We prove that (i) the DEQ
architecture effectively promotes the sparse structure of the output
due to its contraction property when it achieves PMSE-optimality,
and (ii) the combination of DEQ\textquoteright s compressible output
structure and GSURE-based PMSE minimization is sufficient to guarantee
overall MSE performance, approaching the oracle-level accuracy of
supervised training.
\item \textbf{Numerical Verifications:} Numerical experiments demonstrate
the superior performance of the proposed GUDL in terms of reconstruction
accuracy and robustness. We also show that the combination of GSURE
and DEQ can effectively eliminate the performance gap between MSE-based
supervised training and GSURE-based unsupervised training.
\end{itemize}
This paper is organized as follows. In Section \ref{system model},
we present the system model. In Section \ref{methodology}, we introduce
DEQ, GSURE and related implements. The CS-based theoretical analysis,
i.e., Oracle Inequality, is provided in Section \ref{sec:analysis}.
Finally, the simulation results and the conclusion are given in Section
\ref{simulation} and Section \ref{conclusion}, respectively.

Notation: Throughout this paper, $\mathbf{A}^{T}$, $\mathbf{A}^{H}$,
$\mathbf{A}^{\dag}$, $\mathrm{vec}(\mathbf{A})$ and $\text{Tr}(\mathbf{A})$
denote the transpose, Hermitian transpose, Moore-Penrose inverse,
vectorization and trace of matrix $\mathbf{A}$, respectively. $\mathbf{X}\otimes\mathbf{Y}$,
$\mathbf{X}*\mathbf{Y}$, $\mathbf{X}\bullet\mathbf{Y}$, and $\mathbf{X}\odot\mathbf{Y}$
represent the Kronecker product, Khatri--Rao product, transposed
Khatri-Rao product and Hadamard product of matrices $\mathbf{X}$
and $\mathbf{Y}$, respectively. $\Vert\mathbf{a}\Vert_{p}$ and $a(i)$
are the $\ell_{p}$-norm and the $i$-th element of vector $\mathbf{a}$,
respectively. $\left\Vert \mathbf{a}\right\Vert _{\mathbf{C}}$ is
defined as $\mathbf{a}^{T}\mathbf{C}\mathbf{a}$. 

\section{System Model}

\label{system model}

\subsection{Standard Model for LCE}

Consider a real-valued LCE signal model, i.e., recovering a high-dimensional
signal $\mathbf{h}\in\mathbb{R}^{2N}$ from a noisy linear measurement
$\mathbf{y}\in\mathbb{R}^{2M}$ of the form 
\begin{equation}
\mathbf{y}=\mathbf{Ah}+\mathbf{n},\label{CS}
\end{equation}
where $\mathbf{A}\in\mathbb{R}^{2M\times2N}$ represents a measurement
matrix with $2M\ll2N$ and $\mathbf{n}\in\mathbb{R}^{2M}\sim\mathcal{N}\left(\bm{0},\frac{\sigma^{2}}{2}\mathbf{C}\right)$
is Gaussian noise with covariance matrix $\mathbf{C}$. The high-dimensional
signal $\mathbf{h}$ has a sparse structure, i.e., a small number
of coefficients contain a large proportion of the energy. We assume
the measurement matrix $\mathbf{A}$ to be partial orthogonal for
the following reasons: 1) it is easily satisfied in practical systems;
2) the Restricted Isometry Property (RIP) of partially orthogonal
matrix is rigorously established in \cite{haviv2017restricted}, serving
as a foundation for the performance guarantees of CS algorithms; 3)
the orthogonality condition $\mathbf{A}\mathbf{A}^{T}=\mathbf{I}_{2M}$
substantially simplifies algorithmic development and enables theoretical
performance analysis of the proposed method. In the following, we
will explain how LCE problems in several representative scenarios
can be formulated into the above standard form.

\subsection{Sparse Channel Modeling \label{subsec:Application-Examples}}

Consider LCE problems in millimeter-wave (mmWave) massive MIMO systems
with hybrid beamforming (HBF) employed at the base station (BS).
Assume the BS is equipped with $N$ antennas and $S\ll N$ radio frequency
(RF) chains\textit{\emph{, and employs a half-wavelength spaced}}
uniform linear array (ULA). Assume a user is equipped with single
antenna, and $Q$ normalized pilot signals are transmitted from the
user for uplink LCE. Denote $M=QS$. At the $q$-th time slot, hybrid
combining matrix (HCM) $\mathbf{U}_{q}=\mathbf{W}_{q}\mathbf{G}_{q}\in\mathbb{C}^{N\times S}$
is utilized at the BS to measure the channel, where $\mathbf{W}_{q}$
and $\mathbf{G}_{q}$ denote the RF training matrix and baseband training
matrix, respectively. $\mathbf{h}^{u}\in\mathbb{C}^{N}$ denotes the
uplink channel between the user and the BS. Due to the limited number
of RF chains, the number of available measurements is much smaller
than the dimension of large-scale channels, i.e., $M\ll N$. This
motivates the exploitation of inherent structures of mmWave channels
to enhance LCE performance. In a general far-field scenario, the spatial
domain channel $\mathbf{h}^{u}$ can be transformed into a sparse
representation $\mathbf{h}$ via the Discrete Fourier Transform (DFT).
In the following, we illustrate how the sparse channel model applies
to near-field channels and RIS-assisted cascaded channels. 

\subsubsection{Near-Field Channel Model}

With the increase in the array aperture, users tend to be located
within the near-field range of the BS. In the near-field case, the
spherical wave channel $\mathbf{h}^{u}\in\mathbb{C}^{N}$ can be expressed
as
\begin{equation}
\mathbf{h}^{u}=\sum_{l=1}^{L}\alpha_{l}\mathbf{b}_{r}(\theta_{l},r_{l}),\label{near-field channel model-1}
\end{equation}
where $\alpha_{l}$ is the channel gain of the $l$-th path, $\theta_{l}$
and $r_{l}$ denote the angle and distance, respectively, of the user
(for $l=1)$ or scatterer (for $l>1)$ relative to the center of the
antenna array at the BS. $\mathbf{b}_{r}(\cdot)$ is the near-field
array response vector (ARV), given by
\begin{equation}
\begin{array}{l}
\mathbf{b}_{r}(\theta_{l},r_{l})\\
=\frac{1}{\sqrt{N}}\left[e^{-j\frac{2\pi}{\lambda_{c}}(r_{l}^{(1)}-r_{l})},e^{-j\frac{2\pi}{\lambda_{c}}(r_{l}^{(2)}-r_{l})},\cdots,e^{-j\frac{2\pi}{\lambda_{c}}(r_{l}^{(N)}-r_{l})}\right]^{T},
\end{array}\label{spherical wavefront-1}
\end{equation}
with $r_{l}^{n}$ calculated as
\begin{equation}
r_{l}^{(n)}=\sqrt{\left(r_{l}^{(1)}\right)^{2}+\left((n-1)d\right)^{2}-2r_{l}^{(1)}\left(n-1\right)d\cos(\theta_{l})}.
\end{equation}
As in \cite{cui2022channel}, a polar-domain sparsifying matrix $\mathbf{A}_{\mathrm{nf}}\in\mathbb{C}^{N\times NR}$
can be designed to map the near-field channel to a polar domain sparse
representation $\bar{\mathbf{h}}_{\text{nf}}\in\mathbb{C}^{NR}$ :
\begin{equation}
\mathbf{h}^{u}=\mathbf{A}_{\mathrm{nf}}\bar{\mathbf{h}}.\label{eq:polar}
\end{equation}
$\mathbf{A}_{\mathrm{nf}}\in\mathbb{C}^{N\times NR}$ is given by:
\begin{align}
\mathbf{A}_{\mathrm{nf}}= & [\mathbf{b}_{r}(\theta_{0},r_{0}),\cdots,\mathbf{b}_{r}(\theta_{0},r_{R-1}),\cdots\nonumber \\
 & \mathbf{b}_{r}(\theta_{N-1},r_{0}),\cdots,\mathbf{b}_{r}(\theta_{N-1},r_{R-1})],
\end{align}
where cosine values $\cos(\theta_{i}),i\in\{0,\cdots,N-1\}$, are
sampled uniformly within $[-1,1]$, while the distances $r_{s},s\in\{0,\cdots,R-1\}$,
are sampled non-uniformly from the predefined range $\left[r_{\text{min}},r_{\text{max}}\right]$
at the same angle $\theta_{i}$. Due to the limited scatterers in
the environment, the polar domain representation $\bar{\mathbf{h}}\in\mathbb{C}^{NR}$
of the near-field channel is a high-dimensional sparse vector.  The
received baseband signal $\bar{\mathbf{y}}\in\mathbb{C}^{M}$ at the
BS can be written as
\begin{equation}
\bar{\mathbf{y}}=\bar{\mathbf{A}}\bar{\mathbf{h}}+\bar{\mathbf{n}},\label{near-field aggregated baseband signal-1}
\end{equation}
where $\bar{\mathbf{A}}=\mathbf{U\mathbf{\mathbf{A}_{\mathrm{nf}}}}\in\mathbb{C}^{M\times NR}$,
$\mathbf{U}=\left[\mathbf{U}_{1},\mathbf{U}_{2},\cdots,\mathbf{U}_{Q}\right]^{H}\in\mathbb{C}^{M\times N}$,
and $\bar{\mathbf{n}}=\left[\mathbf{n}_{1}^{H}\mathbf{U}_{1},...,\mathbf{n}_{Q}^{H}\mathbf{U}_{Q}\right]^{H}\in\mathbb{C}^{M\times1}$
with $\mathbf{n}_{q}\sim\mathcal{CN}(\mathbf{0},\sigma^{2}\mathbf{I}_{N})$.

\subsubsection{RIS-Assisted Cascaded Channel Model}

In a scenario where an RIS is used to assist the communication, suppose
it is configured as a uniform planar array with $N_{G}=N_{1}\times N_{2}$
reflecting elements. The cascaded channel between the user and the
BS at the $q$-th time slot is given by
\begin{equation}
\mathbf{h}_{q}^{u}=\mathbf{H}_{\text{BR}}^{*}\mathbf{\Phi}_{q}\mathbf{h}_{\text{RU}}^{*}.
\end{equation}
$\mathbf{\Phi}_{q}\in\mathbb{C}^{N_{G}\times N_{G}}$ denotes the
diagonal phase-shifting matrix at the $q$-th time slot. $\mathbf{H}_{\text{BR}}^{*}\in\mathbb{C}^{N\times N_{G}}$
and $\mathbf{h}_{\text{RU}}^{*}\in\mathbb{C}^{N_{G}}$ represent the
channel from the RIS to the BS and the user to the RIS, respectively,
which are given by
\begin{align}
\mathbf{H}_{\text{BR}}^{*} & =\sum_{l=1}^{L}g_{l}\mathbf{a_{\text{BS}}}(\phi_{l})\mathbf{a}_{\text{RIS}}^{H}(\psi_{l}^{\mathrm{azi}},\psi_{l}^{\mathrm{ele}}),\\
\mathbf{h}_{\text{RU}}^{*} & =\sum_{l=1}^{L'}\alpha_{l}\mathbf{a}_{\text{RIS}}(\theta_{l}^{\mathrm{azi}},\theta_{l}^{\mathrm{ele}}),
\end{align}
where $g_{l}$ ($\alpha_{l}$), $\phi_{l}$, $\psi_{l}^{\mathrm{azi}}$
and $\psi_{l}^{\mathrm{ele}}$($\theta_{l}^{\mathrm{azi}}$ and $\theta_{l}^{\mathrm{ele}}$)
denote the complex channel gains, AoA at the BS, azimuth and elevation
AoD (AoA) of the $l$-th path from the RIS to the BS (the user to
the RIS), respectively. $\mathbf{a_{\text{BS}}}(\cdot)$ and $\mathbf{a}_{\text{RIS}}(\cdot,\cdot)$
denote the ARVs at the BS and the RIS arrays, respectively.  The
received baseband signal at the $q$-th time slot is
\begin{equation}
\begin{array}{cl}
\bar{\mathbf{y}}_{q} & =\mathbf{U}_{q}^{H}\mathbf{H}\bm{\phi}_{q}+\mathbf{U}_{q}^{H}\mathbf{n}_{q},\end{array}\label{RIS-assisted q-th baseband signal-1}
\end{equation}
where $\mathbf{H=}\mathbf{H}_{\text{BR}}^{*}\mathrm{diag}\left(\mathbf{h}_{\text{RU}}^{*}\right)\in\mathbb{C}^{N\times N_{G}}$
is the cascaded channel, and $\mathbf{n}_{q}\in\mathbb{C}^{N}\sim\mathcal{CN}(\mathbf{0},\sigma^{2}\mathbf{I}_{N})$
denotes the AWGN. By discretizing the angular domain, the cascaded
channel $\mathbf{H}$ can be transformed into a sparse representation
\cite{wang2020compressed}, i.e.,
\begin{equation}
\begin{array}{cl}
\mathbf{H} & =\mathbf{A}_{\text{BS}}\mathbf{\mathbf{\Lambda}}\mathbf{D}^{T},\end{array}
\end{equation}
where $\mathbf{A}_{\text{BS}}$ is the ARVs at the BS corresponding
to the angular set $\{\phi_{0},\cdots,\phi_{N-1}\}$. $\mathbf{D}=\mathbf{A}_{\text{RIS}}\bullet\mathrm{conj}\left(\mathbf{A}_{\text{RIS}}\right)$,
where $\mathbf{A}_{\text{RIS}}$ is the ARVs at the RIS corresponding
to the angular set $\left\{ (\theta_{0}^{\mathrm{azi}},\theta_{0}^{\mathrm{ele}}),\cdots,(\theta_{N_{G}-1}^{\mathrm{azi}},\theta_{N_{G}-1}^{\mathrm{ele}})\right\} $.
Due to limited scattering of the transmission environment, the angular
domain representation of the cascaded channel $\mathbf{\mathbf{\Lambda}}$
is a sparse matrix. The received signal over $Q$ time slots can be
written as a standard LCE model, i.e.,
\begin{equation}
\begin{array}{cl}
\bar{\mathbf{y}} & =\bar{\mathbf{A}}\bar{\mathbf{h}}+\bar{\mathbf{n}}\end{array},\label{RIS aggregated baseband signal-1}
\end{equation}
where $\bar{\mathbf{y}}=\left[\bar{\mathbf{y}}_{1}^{H},\cdots,\bar{\mathbf{y}}_{Q}^{H}\right]^{H}\in\mathbb{C}^{M}$,
$\bar{\mathbf{n}}=\left[\mathbf{n}_{1}^{H}\mathbf{U}_{1},\cdots,\mathbf{n}_{Q}^{H}\mathbf{U}_{Q}\right]^{H}\in\mathbb{C}^{M}$,
and $\bar{\mathbf{A}}=\left[\left[\begin{array}{c}
\bm{\phi}_{1}^{T}\otimes\mathbf{U}_{1}^{H}\end{array}\right]^{T},\cdots,\left[\bm{\phi}_{Q}^{T}\otimes\mathbf{U}_{Q}^{H}\right]^{T}\right]^{T}\left(\mathbf{D}\otimes\mathbf{A}_{\text{BS}}\right)$. $\bar{\mathbf{h}}=\mathrm{vec}\left(\mathbf{\mathbf{\Lambda}}\right)\in\mathbb{C}^{N\cdot N_{G}^{2}}$
is the sparse high-dimensional vector to be recovered.

Note that, by carefully designing the phase shifting matrix $\mathbf{\Phi}_{q}$
and HCM $\{\mathbf{U}_{q}\}$ as in \cite{liu2018downlink}, we can
guarantee that the resulting measurement matrix $\bar{\mathbf{A}}$
in \eqref{near-field aggregated baseband signal-1} and \eqref{RIS aggregated baseband signal-1}
are both partial unitary. Then, these matrices can be transformed
into an equivalent real-valued form as in \eqref{CS}.

\subsection{Problem Formulation}

We aim to approximate an estimator $\hat{\mathbf{h}}=f(\mathbf{y})$,
that can achieve the minimum mean squared error (MMSE), i.e.,
\begin{equation}
\hat{\mathbf{h}}=\arg\min_{\hat{\mathbf{h}}}\mathbb{E}\left[\left\Vert \hat{\mathbf{h}}-\mathbf{h}\right\Vert _{2}^{2}\right],
\end{equation}
where the expectation is taken over the joint distribution of $\mathbf{h}$
and the observation $\mathbf{y}$. We assume that $\mathbf{h}\sim p_{\mathbf{h}}$,
where $p$ is some distribution that can enforce a sparse structure.
Various classical algorithms have been developed for sparse recovery.
Deterministic algorithms \cite{wright2009sparse,beck2009fast} solve
a convex least absolute shrinkage and selection operator (LASSO) problem
but do not utilize the statistics of the underlying signals. In contrast,
Bayesian methods \cite{tipping2001sparse,donoho2009message,ji2008bayesian}
incorporate sparse priors to design efficient estimators that approximate
the MMSE solution. However, these classical algorithms suffer from
high computational latency, reliance on unknown priors, and sensitivity
to hyperparameters. To alleviate these issues, in the next section,
we introduce the proposed GUDL framework, which leverages a specialized
DEQ to directly learn the fixed point of an iterative algorithm and
GSURE loss function to optimize the performance with only noisy measurements.\textcolor{blue}{}

\section{GSURE-based Unsupervised DEQ for LCE}

\label{methodology}

\subsection{Deep Equilibrium Model}

\begin{figure}[!t]
\subfloat[Diagram of DU\label{fig: Unrolling}]{\begin{centering}
\includegraphics[viewport=230bp 230bp 605bp 365bp,clip,scale=0.38]{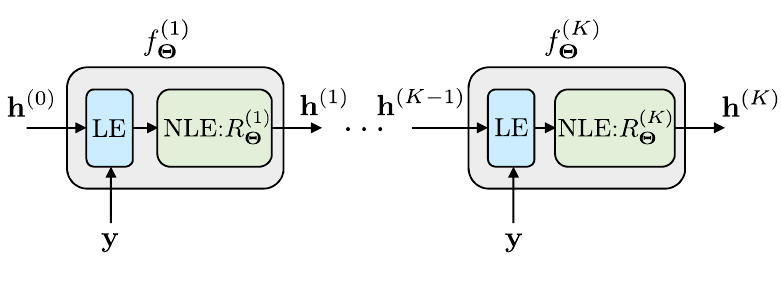}
\par\end{centering}
}\subfloat[Diagram of DEQ\label{fig: Equilibrium}]{\begin{centering}
\includegraphics[viewport=270bp 220bp 520bp 395bp,clip,scale=0.38]{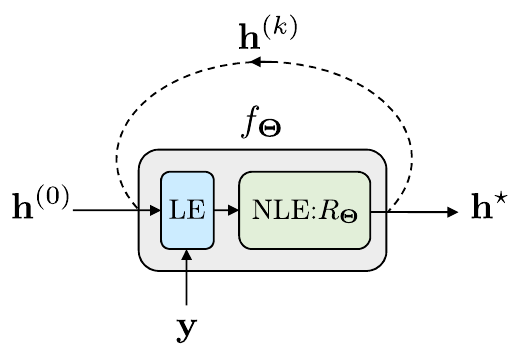}
\par\end{centering}
}\caption{The difference between DU and DEQ.\label{difference_DU_DEQ}}
\end{figure}

The structure of our network is inspired by a standard proximal algorithm,
which recovers the sparse channel $\mathbf{h}$ by solving the following
optimization problem:
\begin{equation}
\hat{\mathbf{h}}=\arg\min_{\mathbf{h}}\frac{1}{2}\Vert\mathbf{y}-\mathbf{Ah}\Vert_{\mathbf{C}^{-1}}^{2}+\tau r_{\mathbf{\Theta}}(\mathbf{h}),\label{optimization}
\end{equation}
where $\tau>0$ is a constant, and $r_{\mathbf{\Theta}}:\mathbb{R}^{2N}\to\mathbb{R}$
is the regularizer with parameters $\mathbf{\Theta}$, designed to
capture the distributional priors of sparse-like vector $\mathbf{h}$.
A proximal gradient descent (PGD) algorithm \cite{parikh2014proximal}
introduces a proximal operator associated with $r_{\mathbf{\Theta}}(\cdot)$
and conducts the following iteration: 
\begin{equation}
\mathbf{h}^{(k)}=\mathrm{prox}_{\tau r_{\mathbf{\Theta}}}\left(\mathbf{h}^{(k-1)}+\eta\mathbf{A}^{T}\mathbf{C}^{-1}(\mathbf{y}-\mathbf{A}\mathbf{h}^{(k-1)})\right),\label{PGD}
\end{equation}
where $\eta>0$ is the stepsize. To improve the algorithmic efficiency,
the proximal operator can be replaced with a neural network to learn
the optimal denoising function directly from the data. Fixing the
number of iterations to $K$, the conventional DU architecture adopts
the following layered structure:
\begin{align}
\text{DU:}\mathbf{h}^{(k)}= & f_{\mathbf{\Theta}}^{(k)}(\mathbf{h}^{(k-1)};\mathbf{y})\nonumber \\
= & R_{\mathbf{\Theta}}^{(k)}\left(\mathbf{h}^{(k-1)}+\eta\mathbf{A}^{T}\mathbf{C}^{-1}(\mathbf{y}-\mathbf{A}\mathbf{h}^{(k-1)})\right),\label{DU}
\end{align}
for $k=1,\cdots,K$, where $R_{\mathbf{\Theta}}^{(k)}$ denotes the
nonlinear estimator (NLE) for layer $k$ and $\mathbf{y}$ is the
input observation. Despite its effectiveness, DU suffers from key
limitations: the layer-wise architecture introduces model instability,
high memory and storage costs, and restricted learning accuracy due
to the limited number of layers. As illustrated in Fig. \ref{difference_DU_DEQ},
DEQ addresses these issues by directly learning the fixed point of
the iterative process with a one-layer neural network. In this sense,
DEQ can be considered as an implicit infinite-depth neural network,
which corresponds to an infinite number of iterations. 

\subsubsection{Model of DEQ}

Let the mapping function of each layer be the same, i.e., $f_{\mathbf{\Theta}}^{(k)}=f_{\mathbf{\Theta}}$
for all $k$. Under this assumption, we have 
\begin{equation}
\mathbf{h}^{(k)}=f_{\mathbf{\Theta}}(\mathbf{h}^{(k-1)};\mathbf{y})\label{weight-tying}
\end{equation}
for $k=1,\cdots,K.$ When $K\to\infty$, we can obtain the fixed-point
solution $\mathbf{h}^{(\infty)}$ of $f_{\mathbf{\mathbf{\Theta}}}$
if it exists. The DEQ directly learns the fixed-point $\mathbf{\mathbf{h}^{\star}}=\mathbf{h}^{(\infty)}$
of $f_{\mathbf{\mathbf{\Theta}}}$ through
\begin{align}
\text{DEQ:}\ \mathbf{h}^{\star}= & f_{\mathbf{\Theta}}(\mathbf{h}^{\star};\mathbf{y})\nonumber \\
= & R_{\mathbf{\Theta}}\left(\mathbf{h}^{\star}+\eta\mathbf{A}^{T}\mathbf{C}^{-1}(\mathbf{y}-\mathbf{A}\mathbf{h}^{\star})\right).\label{DEQ}
\end{align}
In particular, the observation $\mathbf{y}$ is first passed through
a linear estimator (LE), followed by the NLE. The output is recursively
fed back into the network input until convergence is reached, thereby
allowing the DEQ to learn the equilibrium solution $\mathbf{h}^{\star}$.
In what follows, we denote the fixed point of the network $f_{\mathbf{\Theta}}(\cdot;\mathbf{y})$
with parameters $\mathbf{\Theta}$ and input $\mathbf{y}$ as $\mathbf{h}_{\mathbf{\Theta}}^{\star}(\mathbf{y})$,
which may also be abbreviated as $\mathbf{h}_{\mathbf{\Theta}}^{\star}$
or $\mathbf{h}^{\star}$ in some contexts.

\subsubsection{Training of DEQ}

We propose to train the DEQ in an unsupervised manner with the GSURE
loss function, denoted as $\mathcal{L}(\mathbf{h}_{\mathbf{\Theta}}^{\star}(\mathbf{y}),\mathbf{y})$,
which will be elaborated in Section \ref{subsec:Generalized-Stein's-Unbiased}.
During backpropagation, we can calculate the gradient of $\mathcal{L}$
w.r.t. the network parameters $\mathbf{\Theta}$ using implicit differentiation
through the fixed-point $\mathbf{h}^{\star}$. Specifically, the loss
gradient w.r.t $\mathbf{\mathbf{\Theta}}$ is 
\begin{equation}
\frac{\partial\mathcal{L}}{\partial\mathbf{\Theta}}=\frac{\partial\mathbf{h}^{\star}}{\partial\mathbf{\Theta}}^{T}\frac{\partial\mathcal{L}}{\partial\mathbf{h}^{\star}},\label{dl_dtheta}
\end{equation}
where the Jacobian $\frac{\partial\mathbf{h}^{\star}}{\partial\mathbf{\Theta}}$
can be calculated through the fixed-point equation: $\mathbf{h}^{\star}=f_{\mathbf{\Theta}}(\mathbf{\mathbf{\mathbf{h}^{\star}}};\mathbf{y})$,
given by 
\begin{equation}
\frac{\partial\mathbf{h}^{\star}}{\partial\mathbf{\Theta}}=\left(\mathbf{I}_{2N}-\left.\frac{\partial f_{\mathbf{\Theta}}(\mathbf{\mathbf{\mathbf{h}^{\star}}};\mathbf{y})}{\partial\mathbf{h}}\right|_{\mathbf{h}=\mathbf{h}^{\star}}\right)^{-1}\frac{\partial f_{\mathbf{\Theta}}(\mathbf{h}^{\star};\mathbf{y})}{\partial\mathbf{\Theta}}.\label{dh_dtheta}
\end{equation}
Note that the matrix inversion in the Jacobian is still relatively
complicated, especially for large-scale problems. To simplify the
calculation, we adopt the inexact gradient estimate, i.e.,
\begin{equation}
\widehat{\frac{\partial\mathcal{L}}{\partial\mathbf{\Theta}}}\approx\frac{\partial f_{\mathbf{\Theta}}(\mathbf{\mathbf{\mathbf{h}^{\star}}};\mathbf{y})}{\partial\mathbf{\mathbf{\Theta}}}^{T}\frac{\partial\mathcal{L}}{\partial\mathbf{h}^{\star}},\label{approx grad}
\end{equation}
in which the matrix inversion vanishes. Theorem 3.1 in \cite{fung2022jfb}
proves that the approximation of $\widehat{\frac{\partial\mathcal{L}}{\partial\mathbf{\Theta}}}$
in \eqref{approx grad} is a descent direction of $\mathcal{L}(\mathbf{h}_{\mathbf{\Theta}}^{\star}(\mathbf{y}),\mathbf{y})$
with respect to $\mathbf{\Theta}$ under mild conditions. Therefore,
the original calculation-intensive task of backpropagation over an
infinite number of layers has been converted into one layer of calculation
with constant memory.

\subsubsection{Convergence of DEQ}

To ensure the convergence of DEQ, we design the mapping function $f_{\mathbf{\Theta}}(\cdot;\mathbf{y})$
as a contraction mapping.

\begin{definition}[Contraction] $f_{\mathbf{\Theta}}(\cdot;\mathbf{y})$
is a contraction operation if there exists a constant $L\in(0,1)$,
such that 
\begin{equation}
\Vert f_{\mathbf{\Theta}}(\mathbf{x}_{1};\mathbf{y})-f_{\mathbf{\Theta}}(\mathbf{x}_{2};\mathbf{y})\Vert_{2}\le L\cdot\Vert\mathbf{x}_{1}-\mathbf{x}_{2}\Vert_{2}\label{contraction}
\end{equation}
holds for $\forall\mathbf{x}_{1},\mathbf{x}_{2}\in\mathbb{R}^{2N}$.

\end{definition}

\begin{lemma}[Convergence Under Contraction\cite{bauschke10convex}]
Let the mapping function of DEQ $f_{\mathbf{\Theta}}(\cdot;\mathbf{y})$
be a Lipschitz contraction with $L\in(0,1)$. For any initialization
$\mathbf{h}^{(0)}\in\mathbb{R}^{2N}$, the sequence generated by DEQ
update $\{\mathbf{h}^{(k)}\}$ converges to the unique fixed point
$\mathbf{h}^{\star}\in\mathbb{R}^{2N}$ of $f_{\mathbf{\Theta}}(\cdot;\mathbf{y})$
with a linear convergence rate.

\end{lemma}

The Lipschitz constant $L$ of $f_{\mathbf{\Theta}}(\mathbf{\cdot};\mathbf{y})$
can be calculated as $L=L_{1}L_{2}$, where $L_{1}$ is the Lipschitz
constant of the LE, and $L_{2}$ is the Lipschitz constant of the
NLE $R_{\mathbf{\Theta}}(\cdot)$. $L_{1}$ can be controlled to equal
1 via tuning the learning rate $\eta$. $L_{2}$ should satisfy $0<L_{2}<1$,
which can be controlled during training through known methods \cite{gouk2021regularisation}.
This contraction assumption not only guarantees convergence and robustness
to distribution shifts, but also ensures compressibility of DEQ outputs
under GSURE optimality, supporting theoretical guarantees for the
learning performance.

\subsection{Generalized Stein's Unbiased Risk Estimate}

\label{subsec:Generalized-Stein's-Unbiased}

Considering the standard LCE signal model in \eqref{CS},  traditional
SURE assumption, i.e., AWGN observation model \cite{stein1981estimation},
becomes invalid. GSURE \cite{eldar2008generalized} has extended SURE
to more general linear measurement model and probability distributions.
Specifically, considering the exponential family of probability distributions
of $p(\mathbf{y};\mathbf{h})$, the sufficient statistic for estimating
$\mathbf{h}$ from $\mathbf{y}$ is given by $\frac{2}{\sigma^{2}}\mathbf{A}^{T}\mathbf{C}^{-1}\mathbf{y}$,
and any reasonable estimate of $\mathbf{h}$ would be a function of
the sufficient statistic. In this paper, we adopt $\mathbf{u}=\mathbf{A}^{T}\mathbf{C}^{-1}\mathbf{y}$
as network input, which not only facilitates algorithmic derivation
but also simplifies theoretical analysis. Since the measurement matrix
$\mathbf{A}$ is designed to be partial orthogonal, i.e., $\mathbf{A}\mathbf{A}^{T}=\mathbf{I}_{2M}$,
$\mathbf{u}$ lies in the range space $\mathcal{A}=\mathcal{R}(\mathbf{A}^{T})$
of $\mathbf{A}^{T}$, which is a subspace of $\mathbb{R}^{2N}$. Based
on $\mathbf{u}$, we can obtain a reliable estimate of part of $\mathbf{h}$
that lies in $\mathcal{A}$. Denote the estimate of $\mathbf{h}$
as $\hat{\mathbf{h}}=g(\mathbf{u})$. The MSE in estimating $\mathbf{h}$
is given by 
\begin{align}
\mathrm{MSE}= & \mathbb{E}\Vert\hat{\mathbf{h}}-\mathbf{h}\Vert_{2}^{2}\nonumber \\
= & \underbrace{\mathbb{E}\Vert\mathbf{P}(\hat{\mathbf{h}}-\mathbf{h})\Vert_{2}^{2}}_{\text{PMSE}}+\underbrace{\mathbb{E}\Vert(\mathbf{I}_{2N}-\mathbf{P})(\hat{\mathbf{h}}-\mathbf{h})\Vert_{2}^{2}}_{\mathrm{RMSE}},\label{MSE}
\end{align}
where $\mathbf{P}=\mathbf{A}^{T}(\mathbf{A}\mathbf{A}^{T})^{-1}\mathbf{A}=\mathbf{A}^{T}\mathbf{A}$
denotes the orthogonal projection onto $\mathcal{A}$. The first term
in \eqref{MSE} represents the PMSE, which measures the estimation
error within space $\mathcal{A}$, while the second term accounts
for the residual MSE (RMSE), which does not lie in subspace $\mathcal{A}$.
If $\hat{\mathbf{h}}$ lies in $\mathcal{A}$, the second term is
irrelevant to $\hat{\mathbf{h}}$ and it is sufficient to minimize
the first term to obtain an optimal MMSE estimate of $\mathbf{h}$.
If $\mathbf{\hat{\mathbf{h}}}$ doesn't belong to $\mathcal{A}$,
to obtain the MMSE estimate, the PMSE and RMSE need to be minimized
simultaneously. GSURE provides an unbiased estimate of PMSE, which
is given by
\begin{align}
 & \mathrm{PMSE}=\mathbb{E}\Vert\mathbf{P}\hat{\mathbf{h}}-\mathbf{Ph}\Vert_{2}^{2}=\mathbb{E}\Vert\mathbf{P}g(\mathbf{u})-\mathbf{Ph}\Vert_{2}^{2}\nonumber \\
= & \underbrace{\mathbb{E}\left[\left\Vert \mathbf{P}g(\mathbf{u})-\mathbf{A}^{T}\mathbf{C}\mathbf{A}\mathbf{u}\right\Vert _{2}^{2}+\sigma^{2}\mathrm{Tr}\left(\mathbf{\mathbf{P}}\frac{\partial g(\mathbf{u})}{\partial\mathbf{u}}-\mathbf{C}\right)\right]}_{\mathrm{\mathrm{GSURE}}\left(g(\mathbf{u})\right)}.\label{PMSE}
\end{align}

It can be observed that $\mathrm{\mathrm{GSURE}}\left(g(\mathbf{u})\right)$
in \eqref{PMSE} provides an unbiased estimate of PMSE without requiring
the knowledge of ground truth $\mathbf{h}$. This allows GSURE to
be employed as an unsupervised loss function. Specifically, the estimator
$g(\mathbf{u})$ in \eqref{PMSE} is instantiated by the DEQ model
$f_{\mathbf{\Theta}}(\cdot;\mathbf{u})$ given in \eqref{DEQ}, by
replacing $\mathbf{A}^{T}\mathbf{C}^{-1}\mathbf{y}$ with $\mathbf{u}$.
 In the following, we define the GSURE-optimal estimate of DEQ.

\begin{definition}[GSURE-Optimal Estimate of DEQ] The GSURE-optimal
estimate of DEQ is defined as the fixed-point of DEQ obtained by minimizing
the GSURE loss function in \eqref{PMSE}, i.e., $\mathbf{h}_{\mathrm{G}}^{\star}=f_{\mathbf{\Theta}_{\mathrm{G}}^{\star}}(\mathbf{h}_{\mathrm{G}}^{\star};\mathbf{u})$,
where $\mathbf{\Theta}_{\mathrm{G}}^{\star}$ is given by
\begin{equation}
\mathbf{\Theta}_{\mathrm{G}}^{\star}=\arg\min_{\mathbf{\Theta}}\mathrm{GSURE}\left(f_{\mathbf{\Theta}}(\cdot;\mathbf{u})\right).\label{Theta_G}
\end{equation}
Let $\mathbf{h}$ denote the ground-truth channel. The MSE and normalized
MSE (NMSE) of GUSRE-optimal solution are defined as 
\begin{equation}
\mathrm{MSE}_{\mathrm{G}}=\mathbb{E}\left[\Vert\mathbf{h}_{\mathrm{G}}^{\star}-\mathbf{h}\Vert_{2}^{2}\right],\mathrm{NMSE}_{\mathrm{G}}=\frac{\mathrm{MSE}_{\mathrm{G}}}{\mathbb{E}\left[\left\Vert \mathbf{h}\right\Vert _{2}^{2}\right]}.
\end{equation}
\end{definition}

After the convergence of DEQ, the fixed-point $\mathbf{h}_{\mathrm{G}}^{\star}$
can be considered as the point that minimizes the PMSE of the LCE.
However, as shown in \eqref{MSE}, the MMSE estimator and the minimum
PMSE estimator are not equivalent. In Section \ref{sec:analysis},
we will show that when large-scale channel exhibit a sparse structure,
training the DEQ network with the GSURE loss function can achieve
the performance close to that of the MMSE estimator.

\section{Performance Guarantees of GUDL\label{sec:analysis}}

In this section, we provide theoretical guarantees that the proposed
GUDL can learn the MMSE estimate of DEQ. Before proceeding with our
theoretical results, we first introduce some key definitions and assumptions.

\begin{definition}[MSE-Optimal (Oracle) Estimate of DEQ]The MSE-optimal
(oracle) estimate of DEQ $\mathbf{h}_{\text{ora}}^{\star}$ is defined
as the fixed-point of DEQ obtained by minimizing the MSE of the estimate,
i.e., $\mathbf{h}_{\mathrm{ora}}^{\star}=f_{\mathbf{\Theta}_{\mathrm{ora}}^{\star}}(\mathbf{h}_{\mathrm{ora}}^{\star};\mathbf{u})$
with $\mathbf{\Theta}_{\mathrm{ora}}^{\star}$ given by
\begin{equation}
\mathbf{\Theta}_{\mathrm{ora}}^{\star}=\arg\min_{\mathbf{\Theta}}\mathbb{E}\left[\Vert f_{\mathbf{\Theta}}(\cdot;\mathbf{u})-\mathbf{h}\Vert_{2}^{2}\right].\label{Theta_MSE}
\end{equation}
The resultin\textcolor{black}{g MSE }of DEQ (oracle MSE) and NMSE
are defined as 
\begin{equation}
\mathrm{MSE}_{\mathrm{ora}}=\mathbb{E}\left[\Vert\mathbf{h}_{\mathrm{ora}}^{\star}-\mathbf{h}\Vert_{2}^{2}\right],\mathrm{NMSE}_{\mathrm{ora}}=\frac{\mathrm{MSE}_{\mathrm{ora}}}{\mathbb{E}\left[\left\Vert \mathbf{h}\right\Vert _{2}^{2}\right]}.
\end{equation}

\end{definition}

\begin{assumption}[Sparsity of Ground Truth]\label{k_sparse} Assume
the real-valued ground-truth channel data $\mathbf{h}\in\mathbb{R}^{2N}$
is at most $2k$-sparse, i.e., $\Vert\mathbf{h}\Vert_{0}\le2k\ll2N.$

\end{assumption}

Assumption \ref{k_sparse} states that the ground-truth channel is
compressible when transformed into the standard model in \eqref{CS},
which has been elaborated in Subsection \ref{subsec:Application-Examples}
through several classical examples.

\begin{assumption}[Consistency of Estimation Errors]\label{good_pmse}
Assume there exist constant $\beta\in(0,1)$ and $\omega\in(0,1)$,
such that for any $\mathbf{h}$, the projected error of $\mathbf{h}_{\mathrm{G}}^{\star}$
and the error of $\mathbf{h}_{\mathrm{ora}}^{\star}$ satisfy the
following conditions:
\begin{equation}
\mathbf{\Vert P}\mathbf{h}_{\mathrm{G}}^{\star}-\mathbf{Ph}\Vert_{2}\le\beta\cdot\Vert\mathbf{Ph}\Vert_{2},~\Vert\mathbf{h}_{\mathrm{ora}}^{\star}-\mathbf{h}\Vert_{2}\le\omega\cdot\Vert\mathbf{h}\Vert_{2}.\label{beta}
\end{equation}

\end{assumption}

Assumption \ref{good_pmse} suggests that the normalized projected
error and the normalized oracle error of $\mathbf{h}_{\mathrm{G}}^{\star}$
and $\mathbf{h}_{\mathrm{ora}}^{\star}$ are bounded above by small
values. Based on the definitions of $\mathbf{h}_{\mathrm{G}}^{\star}$
and $\mathbf{h}_{\mathrm{ora}}^{\star}$, the above assumption is
achievable for a neural network with good representation ability,
provided the SNR is not extremely small. Our simulations in Section
\ref{verify_Oracle} verify that Assumption \ref{good_pmse} can be
easily satisfied.

\begin{assumption}[Zero-Input Zero-Response of DEQ]\label{zero_output}
Assume the NLE part $R_{\mathbf{\Theta}}$ has zero response for zero
input, i.e.,
\begin{equation}
R_{\mathbf{\Theta}}(\bm{0})=\bm{0}.
\end{equation}
\end{assumption}

Assumption \ref{zero_output} ensures that the NLE function does not
introduce bias when no input is present. This can be achieved by setting
the bias of all network layers to zero. This property simplifies theoretical
analysis with minimal impact on performance and may even slightly
improve it, as verified in Section \ref{verify_Oracle}.

\begin{assumption}[Instantaneous SNR Lower Bound]\label{psnr bound}
Assume there exists a constant $\xi>0$ such that, with high probability,
the following instantaneous SNR constraint holds:
\[
\left\Vert \mathbf{Ah}\right\Vert _{2}^{2}/\left\Vert \mathbf{n}\right\Vert _{2}^{2}\geq\xi^{-2}.
\]

\end{assumption}

Since $\|\mathbf{n}\|_{2}^{2}$ is the squared norm of a sub-Gaussian
random vector, the Hanson-Wright concentration inequality \cite{rudelson2013hanson}
yields that a probabilistic upper bound exists for $\|\mathbf{n}\|_{2}^{2}$.
Under this condition, Assumption \ref{psnr bound} reduces to a constraint
on the signal power $\left\Vert \mathbf{Ah}\right\Vert _{2}^{2}$.
The quantity $\xi^{-2}$ characterizes the worse-case SNR level of
the system: a smaller $\xi$ indicates a higher SNR (i.e., better
signal quality), while a larger $\xi$ corresponds to a lower SNR
(i.e., degraded signal quality).

\subsection{Main Result}

We now present the theoretical performance guarantee of the proposed
GUDL in the following main theorem.
\begin{thm}[Oracle Inequality]
\label{oracle inequality} Assume the DEQ defined in equation \eqref{DEQ}
includes a 1-Lipschitz LE and an $L$-Lipschitz contractive NLE $R_{\mathbf{\Theta}}$.
Let $\gamma=\eta L/(1-L)$. Assume the measurement matrix $\mathbf{A}$
satisfies $\Vert\mathbf{A}\Vert_{\infty}\le\zeta/\sqrt{2N}$, where
$\zeta>0$ is a constant. Under Assumptions \ref{k_sparse}-\ref{psnr bound},
there exists a constant $C>0$ such that if
\begin{equation}
M\ge\frac{C\zeta^{2}}{\delta^{2}}\cdot T\cdot\log^{2}\left(\frac{2T}{\delta}\right)\cdot\log^{2}\left(\frac{1}{\delta}\right)\cdot\log\left(2N\right),\label{eq:RIP req}
\end{equation}
then with high probability, the matrix $\sqrt{\frac{N}{M}}\mathbf{A}$
satisfies the RIP of order $2T$ with constant $\delta_{2T}\leq\delta,$
where $T=s+2k,$ and 
\begin{equation}
s=\max\left\{ \left\lceil s_{\mathrm{G}}\right\rceil ,\left\lceil s_{\mathrm{ora}}\right\rceil \right\} 
\end{equation}
with 
\begin{align}
s_{\mathrm{G}}= & \frac{1+\left[\sqrt{2k}+(2N-2k)\cdot2k\right]\cdot\delta}{(1-\beta)(1-\delta_{2k})}\sqrt{2k}\nonumber \\
 & +{\displaystyle \sqrt{2N}\left(\frac{\beta}{1-\beta}+\sqrt{1-\frac{(1-\beta)^{2}}{\gamma^{2}\|\mathbf{C}^{-1}\|_{2}^{2}(\beta+{\displaystyle \xi})^{2}}}\right),}\label{s_G}
\end{align}
\begin{equation}
s_{\mathrm{ora}}=\frac{1}{1-\omega}\sqrt{2k}+\sqrt{2N-2k}\cdot\frac{\omega}{1-\omega}.\label{s_ora}
\end{equation}
Furthermore, under the above conditions, the following oracle inequality
holds with high probability:
\begin{equation}
\left|\mathrm{NMSE}_{\mathrm{G}}-\mathrm{NMSE}_{\mathrm{ora}}\right|\le\left(\epsilon_{1}\beta^{2}+\epsilon_{2}\gamma^{2}g(s)\right)\delta(1+\delta),
\end{equation}
where the constants $\epsilon_{1}$ and $\epsilon_{2}$ are given
by
\begin{gather}
\epsilon_{1}=\frac{2\left(1+\frac{1}{\sqrt{\rho}}\right)}{1-\left(1+\frac{1}{\sqrt{\rho}}\right)^{2}\delta^{2}},\\
\epsilon_{2}=\frac{4\kappa\cdot(1-\rho)}{\rho}\frac{1+\xi^{2}}{1-\left(1+\frac{1}{\sqrt{\rho}}\right)^{2}\delta^{2}}
\end{gather}
with $\rho=\frac{T}{2N},\eta=\frac{M}{N}$, and the function $g(\cdot)$
is given by \eqref{eq: gk}.
\end{thm}
Theorem \ref{oracle inequality} shows that the error between GSURE-
and MSE-optimal DEQ is dominated by the $2T$-order RIP constant $\delta$,
which satisfies $\delta=\mathcal{O}\left(\sqrt{T\log(2T)\log(2N)/M}\right)$.
Here, $T$ relates to the sparsity level $s+2k$, with $s$ denoting
the DEQ output sparsity. As $T\log(2T)\log(2N)/M\to0$, the bound
vanishes, and GUDL approaches oracle performance. The bound also involves
the projection error bound $\beta$, contraction factor $\gamma$,
and the function $g$, which characterize the residual energy from
inexact sparsity of DEQ output. Smaller values of $\beta$, $\gamma$
and $g$ help to tighten the performance gap. The scaling constants
$\epsilon_{1}$ and $\epsilon_{2}$ depend on the sparsity ratio $\rho$
of the error vector, the sampling ratio $\eta$, the SNR-related parameter
$\xi$, and the RIP constant $\delta$. Note that the value of RIP
constant $\delta$ relies on the values of both $\rho$ and $\eta$,
as indicated in \eqref{eq:RIP req}, where a smaller $\rho$ and a
larger $\eta$ generally lead to a smaller $\delta$, which in turn
helps to reduce $\epsilon_{1}$ and $\epsilon_{2}$. However, since
$\epsilon_{1}$ and $\epsilon_{2}$ also contain terms that directly
depend on $\rho$ and $\eta$, their direct contributions may partially
counteract the beneficial effects of a reduced RIP constant. This
coupling complicates a straightforward analysis of the individual
effects of $\rho$ and $\eta$. In contrast, improved signal quality,
reflected by a smaller $\xi$, consistently contributes to a smaller
$\epsilon_{2}$, thus tightening the error bound.

\subsection{Analysis of Sparse Growth Function}

Before proving the main theorem, we introduce the sparse growth function
(SGF), which serves as a fundamental metric in our analysis. This
function quantifies the proportion of residual energy remaining after
a vector is approximated by a sparse representation under a specified
sparsity measure. In particular, it captures how the the energy of
a signal decays beyond its significant components, thereby providing
a quantitative characterization of the signal's compressibility. In
the following, we provide the formal definition of the SGF when $\ell_{1/2}$
is utilized to measure the sparsity.

\begin{definition}[SGF under $\ell_{1/2}$ Sparsity] For a vector
$\boldsymbol{\alpha}\in\mathbb{R}^{2N}$, the $\ell_{1/2}$ sparsity
measure is defined as:
\begin{equation}
\ell_{1/2}(\bm{\alpha})=\frac{\Vert\bm{\alpha}\Vert_{1}}{\Vert\bm{\alpha}\Vert_{2}}.
\end{equation}
Given a positive integer $\kappa\le2N,$ the $\kappa$-sparse approximation
of $\boldsymbol{\alpha}$, denoted by $\bm{\alpha}_{\kappa}$, is
obtained by keeping the $\kappa$ largest entries of $\boldsymbol{\alpha}$
and setting the remaining entries to zero. That is: 
\[
\bm{\alpha}_{\kappa}\in\arg\min_{\left\Vert \boldsymbol{x}\right\Vert _{0}\leq\kappa}\left\Vert \boldsymbol{\alpha}-\boldsymbol{x}\right\Vert _{2}^{2}.
\]
The residual vector is defined as $\bm{\alpha}_{\kappa^{c}}=\boldsymbol{\alpha}-\bm{\alpha}_{\kappa}$.
Then, the SGF under $\ell_{1/2}$ sparsity measure is defined as:
\begin{equation}
B_{\ell_{1/2}}(\kappa,\sqrt{\kappa})=\sup_{\begin{array}{c}
\ell_{1/2}(\bm{\alpha})\le\sqrt{\kappa}\end{array}}\frac{\Vert\bm{\alpha}_{\kappa^{c}}\Vert_{2}^{2}}{\Vert\bm{\alpha}\Vert_{2}^{2}}.
\end{equation}

\end{definition}

A sparsity metric should measure the extent to which a small number
of coefficients contain a large proportion of the energy \cite{hurley2009comparing}.
Here we adopt the ratio of $\ell_{1}$-norm and $\ell_{2}$-norm as
the metric. Many works \cite{xu2021analysis} point out that the $\ell_{1/2}$
of a vector can be used to measure sparsity. When the $\ell_{1/2}$
function is used as the optimization objective in compressed sensing
recovery, the solution tends to be sparser and even has better robustness.
The SGF $B_{\ell_{1/2}}(\kappa,\sqrt{\kappa})$ quantifies the maximum
proportion of residual energy (outside the top $\kappa$ entries)
relative to the total energy, over all vectors whose $\ell_{1/2}$
sparsity does not exceed $\sqrt{\kappa}$. It serves as a useful tool
for characterizing the approximation error of compressible signals
under a sparsity constraint. Intuitively, a dense vector $\mathbf{x}\in\mathbb{R}^{2N}$
has $\Vert\mathbf{x}\Vert_{1}\le\sqrt{2N}\Vert\mathbf{x}\Vert_{2}$.
If $\mathbf{x}$ is exactly $\kappa$-sparse, there is $\Vert\mathbf{x}\Vert_{1}\le\sqrt{\kappa}\Vert\mathbf{x}\Vert_{2}$,
and $B_{\ell_{1/2}}(\kappa,\sqrt{\kappa})=0.$ Therefore, a lower
$\ell_{1/2}$ sparsity measure implies that the signal energy is more
concentrated in fewer components, resulting in a reduced residual
energy and a smaller SGF $B_{\ell_{1/2}}$. In the following theorem,
we quantify the upper bound of SGF under $\ell_{1/2}$ sparsity measure.
\begin{thm}[SGF Upper Bound]
\label{SGF Upper Bound} Given a real-valued vector $\boldsymbol{\alpha}\in\mathbb{R}^{2N}$
and a positive integer $\kappa\le2N,$ the SGF of $\boldsymbol{\alpha}$
under the $\ell_{1/2}$ sparsity constraint admits the following upper
bound:
\begin{equation}
B_{\ell_{1/2}}(\kappa,\sqrt{\kappa})\le g(\kappa),\label{B_constant}
\end{equation}
where the bounding function $g(\kappa)$ is given by 
\begin{equation}
g(\kappa)=\sup_{x\in(0,2N-\kappa)}\frac{x\cdot\tilde{\rho}^{2}(\kappa,x)}{1+(x+\kappa-1)\cdot\tilde{\rho}^{2}(\kappa,x)}\label{eq: gk}
\end{equation}
with
\begin{gather}
\tilde{\rho}(\kappa,x)=\left(\sqrt{\frac{x\cdot\kappa}{x+\kappa-1}}-1\right)/(x-1).
\end{gather}
\end{thm}
The proof is provided in Appendix \ref{proof of SGF Upper Bound}.
Through optimality conditions, we can also derive the closed-form
expression for $g(\kappa)$, denoted as $\tilde{g}(\kappa)$, as shown
in \eqref{eq:closed form gk}. This result plays a critical role
in the proof of our main theorem. The main challenge when analyzing
the performance of GUDL is that the output of the converged DEQ can
be proven to be compressible rather than exact sparse. As a result,
a portion of the energy is distributed outside the dominant components,
leading to a residual that complicates the recovery error analysis.
Fortunately, SGF quantitatively characterizes the proportion of the
residual energy, enabling a precise analysis of the final recovery
error using the RIP theory.

\subsection{Roadmap to Main Theorem}

The first key result needed for the proof of Theorem \ref{oracle inequality}
is to show that the output of the converged DEQ exhibits compressibility,
and its level can be effectively quantified using the $\ell_{1/2}$
sparsity metric.
\begin{thm}
[Bounded $\ell_{1/2}$ Sparsity of $\mathbf{h}_{\mathrm{G}}^{\star}$
and $\mathbf{h}_{\mathrm{ora}}^{\star}$]\label{bounded l_1/2} Assume
the DEQ defined in equation \eqref{DEQ} includes a 1-Lipschitz LE
and an $L$-Lipschitz contractive NLE $R_{\mathbf{\Theta}}$. Let
$\gamma=\eta L/(1-L)$. Let $\delta_{2k}$ denote the $2k$-order
RIP constant of $\sqrt{\frac{N}{M}}\mathbf{A}$. Under Assumptions\textcolor{blue}{{}
\ref{k_sparse}-\ref{psnr bound}}, it holds that 
\begin{align}
\ell_{1/2}(\mathbf{h}_{\mathrm{G}}^{\star})\le & \frac{1+\left[\sqrt{2k}\cdot\delta_{2k}+4k(N-k)\delta_{2}\right]}{(1-\beta)(1-\delta_{2k})}\sqrt{2k}\nonumber \\
 & +\sqrt{2N}\left(\frac{\beta}{1-\beta}+\sqrt{1-\frac{(1-\beta)^{2}}{\gamma^{2}\|\mathbf{C}^{-1}\|_{2}^{2}(\beta+{\displaystyle \xi})^{2}}}\right)\label{Sparsity Measure_G}
\end{align}
and 
\begin{equation}
\ell_{1/2}(\mathbf{h}_{\mathrm{ora}}^{\star})\leq\frac{1}{1-\omega}\sqrt{2k}+\sqrt{2N-2k}\cdot\frac{\omega}{1-\omega}.\label{Sparsity Measure ora}
\end{equation}
\end{thm}
The proof is given in Appendix \ref{proof_bounded l_1/2}. Theorem
\ref{bounded l_1/2} states that both the GSURE-optimal and MSE-optimal
outputs of DEQ exhibit inherent sparse structures. This can be attributed
to the contraction property of the DEQ network, as well as the effectiveness
of the employed loss functions, i.e., GSURE and MSE, in promoting
compressible solutions. 1) For the GSURE-optimal output $\mathbf{h}_{\mathrm{G}}^{\star}$,
the upper bound comprises two terms. The first term approaches the
ground-truth sparsity level (with maximum $\ell_{1/2}$ level of $\sqrt{2k}$)
when the RIP constant $\delta$ and the subspace recovery error $\beta$
are small. The second term vanishes as $\beta\rightarrow0,$ and $\gamma\rightarrow\frac{1}{\xi\cdot||\mathbf{C}^{-1}\|_{2}}$,
where $\frac{1}{\xi\cdot||\mathbf{C}^{-1}\|_{2}}$ serves as a lower
bound on the system SNR when the noise is normalized. Hence, under
favorable conditions, the GSURE-optimal output closely matches the
ground-truth sparsity pattern, with negligible residual-induced discrepancy.
2) For the MSE-optimal output $\mathbf{h}_{\mathrm{ora}}^{\star}$,
both terms in the upper bound depends on the oracle MSE bound $\omega.$
As $\omega\rightarrow0$, the bound approaches the sparsity level
of the ground truth, implying perfect sparsity alignment. Comparing
the two results, $\mathbf{h}_{\mathrm{ora}}^{\star}$ achieves a sparsity
pattern that more closely aligns with the ground-truth channel $\mathbf{h}$,
compared to $\mathbf{h}_{\mathrm{G}}^{\star}$, which exhibits a larger
sparsity discrepancy due to residual energy leakage. 

The third main result required for the proof of Theorem \ref{oracle inequality}
is to exploit the sparsity structures of the converged DEQ output
to quantify the recovery qualities of $\mathbf{h}_{\mathrm{G}}^{\star}$
and $\mathbf{h}_{\mathrm{ora}}^{\star}$ based on the RIP theory.
\begin{thm}[MSE Bound of $\mathbf{h}_{\mathrm{G}}^{\star}$ and $\mathbf{h}_{\mathrm{ora}}^{\star}$]
\label{MSE bound of h_G and h_ora} Under Assumption \ref{k_sparse},
define $\mathcal{T}=\mathrm{supp}(\mathbf{h}_{\mathrm{G},s}^{\star})\cup\mathrm{supp}(\mathbf{h})$,
where $\mathbf{h}_{\mathrm{G},s}^{\star}$ denotes the best $s$-sparse
approximation of $\mathbf{h}_{\mathrm{G}}^{\star}$. Let $\mathbf{h}_{\mathrm{G},\mathcal{T}}^{\star}$
be the vector that keeps the values of $\mathbf{h}_{\mathrm{G}}^{\star}$
on $\mathcal{T}$ and sets the remaining entries to zero. Define $\mathbf{h}_{\mathrm{G},\mathcal{T}^{c}}^{\star}=\mathbf{h}_{\mathrm{G}}^{\star}-\mathbf{h}_{\mathrm{G},\mathcal{T}}^{\star}.$
Similarly, define $\mathcal{R}=\mathrm{supp}(\mathbf{h}_{\mathrm{ora},s}^{\star})\cup\mathrm{supp}(\mathbf{h})$,
and let $\mathbf{h}_{\mathrm{ora},\mathcal{R}}^{\star}$ denote the
vector that retains the entries of $\mathbf{h}_{\mathrm{ora}}^{\star}$
on $\mathcal{R}$, with $\mathbf{h}_{\mathrm{ora},\mathcal{R}^{c}}^{\star}=\mathbf{h}_{\mathrm{ora}}^{\star}-\mathbf{h}_{\mathrm{ora},\mathcal{R}}^{\star}$.
Letting $T=s+2k$, we have $\left\Vert \mathbf{h}_{\mathrm{G},\mathcal{T}}^{\star}\right\Vert _{0}\leq T$
and $\left\Vert \mathbf{h}_{\mathrm{ora},\mathcal{R}}^{\star}\right\Vert _{0}\leq T$.
Assume $\Vert\mathbf{A}\Vert_{\infty}\le\zeta/\sqrt{2N}$. According
to \cite{haviv2017restricted}, there exists a constant $C$ such
that if
\begin{equation}
M\ge\frac{C\zeta^{2}}{\delta^{2}}\cdot T\cdot\log^{2}\left(\frac{2T}{\delta}\right)\cdot\log^{2}\left(\frac{1}{\delta}\right)\cdot\log\left(2N\right),\label{T_le_N}
\end{equation}
then the $2T$-order RIP constant of the matrix $\sqrt{\frac{N}{M}}\mathbf{A}$
satisfies $\delta_{2T}\le\delta$ with high probability. Define $\epsilon=\delta+\sqrt{\frac{2N}{T}}\delta.$
Then, the following inequalities hold with high probability:
\begin{equation}
\left(1-\epsilon\right)\mathrm{MSE}_{\mathrm{G}}\le{\displaystyle \frac{N}{M}}\mathrm{PMSE}(\mathbf{h}_{\mathrm{G}}^{\star})+\delta\cdot{\displaystyle \frac{2N-T}{T}}\mathbb{E}\Vert\mathbf{h}_{\mathrm{G},\mathcal{T}^{c}}^{\star}\Vert_{2}^{2},\label{MSE_le_GSURE}
\end{equation}
\begin{equation}
\left(1+\epsilon\right)\mathrm{MSE}_{\mathrm{ora}}\ge{\displaystyle \frac{N}{M}}\mathrm{\mathrm{PMSE}}(\mathbf{h}_{\mathrm{ora}}^{\star})-\delta\cdot{\displaystyle \frac{2N-T}{T}}\mathbb{E}\Vert\mathbf{h}_{\mathrm{ora},\mathcal{R}^{c}}^{\star}\Vert_{2}^{2}.\label{MSE_ge_GSURE}
\end{equation}
\end{thm}
The proof of Theorem \ref{MSE bound of h_G and h_ora} is given in
Appendix \ref{proof of MSE bound of h_G and h_ora}. Theorem \ref{MSE bound of h_G and h_ora}
states that the MSE of $\mathbf{h}_{\mathrm{G}}^{\star}$ is upper
bounded by its PMSE plus an additional residual error. The residual
error arises from the energy leakage due to the non-exact sparsity
of the GSURE-optimal output and is further influenced by the RIP constant.
Similarly, the MSE of $\mathbf{h}_{\mathrm{ora}}^{\star}$ is lower
bounded by its PMSE minus an additional residual error, which also
depends on the RIP constant and the residual energy. These residual
terms can be quantified using the SGF upper bound. Finally, we arrive
at our main Theorem \ref{oracle inequality}, whose proof, given in
Appendix \ref{proof of oracle inequality}, combines the results in
Theorem \ref{MSE bound of h_G and h_ora} to analyze the gap between
$\mathrm{NMSE}_{\mathrm{G}}$ and $\mathrm{NMSE}_{\mathrm{ora}}$.

\section{Simulations}

\label{simulation} In this section, we provide extensive simulation
results to validate the soundness and correctness of the theoretical
analysis. Furthermore, we evaluate the practical performance of the
proposed GUDL on LCE problem using public datasets, demonstrating
its effectiveness in terms of both recovery accuracy and robustness.

\subsection{Simulation Setups}

\begin{figure}[!t]
\begin{centering}
\includegraphics[viewport=65bp 210bp 772bp 440bp,clip,scale=0.35]{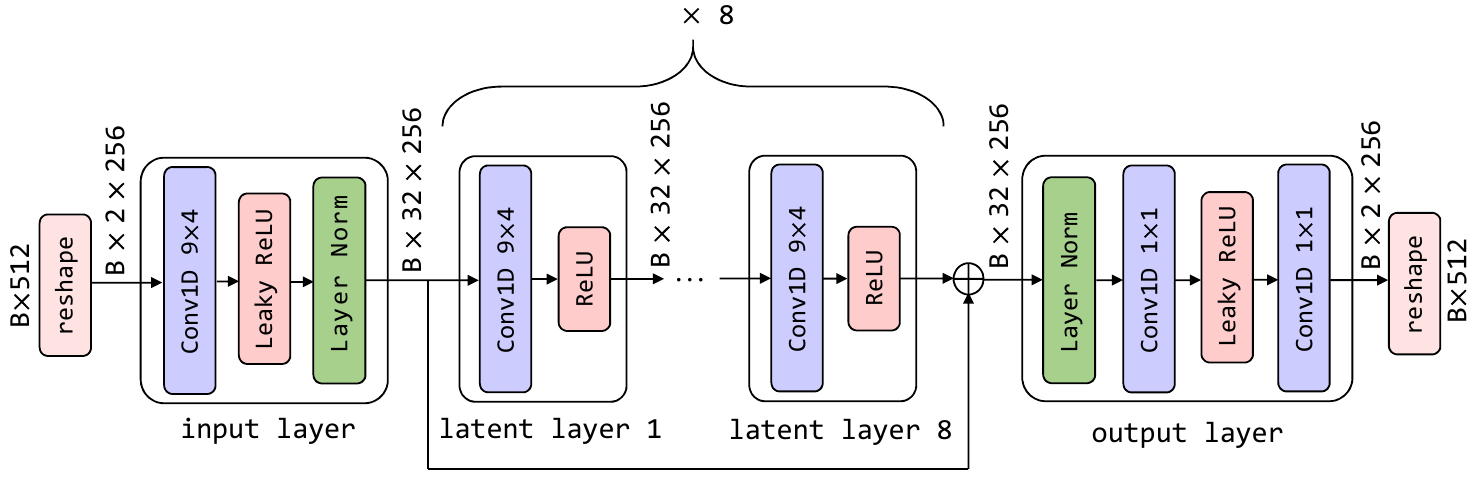}
\par\end{centering}
\centering{}\caption{The structure of the nonlinear part in DEQ. Convolutional layer parameters
are denoted as: kernel size $\times$ padding. Feature size is denoted
as: batch size $\times$ channels $\times$ dimension. \label{R_Theta}}
\end{figure}

The DeepMIMO dataset \cite{alkhateeb2019deepmimo}, a public dataset
for deep learning applications in mmWave and massive MIMO systems,
is adopted for performance evaluation, unless otherwise specified.
The main system parameters are listed in Table \ref{Key Parameters}.
The structure of NLE is shown in Fig. \ref{R_Theta}. We consider
the following baselines:
\begin{itemize}
\item Classical optimization-based algorithms, including OMP \cite{pati1993orthogonal},
AMP \cite{wei2020deep} and OAMP \cite{ma2014turbo};
\item Supervised learning methods, including DNN-NMSE, where a fully connected
DNN with 4 hidden layers (each with width 2048, ReLU activation) is
trained to minimize the NMSE, LDGEC-NMSE \cite{he2022beamspace},
where an unrolled LDGEC network is trained with NMSE loss, DEQ-PMSE,
where the proposed DEQ network is trained using the supervised PMSE
loss, and DEQ-NMSE, where the proposed DEQ network is trained under
the supervised NMSE criterion;
\item Unsupervised learning methods, including DNN-GSURE, where a DNN is
trained using the proposed GSURE loss, LDGEC-SURE, where an unrolled
LDGEC network is trained by minimizing the SURE loss, and DEQ-GSURE,
which stands for the proposed GUDL framework.
\end{itemize}
The proposed DEQ-GSURE is trained using Adam optimizer for 100 epochs.
The batchsize is 128 and the initial learning rate is 0.001, which
decays by half every 30 epochs. Note that in the UCL baselines, the
training dataset only contains the received measurements. The LCE
performance is measured using $\mathrm{NMSE}=\mathbb{E}\big[\Vert\hat{\mathbf{h}}-\mathbf{h}\Vert_{2}^{2}\big]/\mathbb{E}\Vert\mathbf{h}\Vert_{2}^{2}$.

\begin{table}[!t]
\centering{}\caption{KEY SYSTEM PARAMETERS\label{Key Parameters}}
\begin{tabular}{c|c}
\hline 
\textbf{Parameters} & \textbf{Value}\tabularnewline
\hline 
Number of BS Antennas & $N=256$\tabularnewline
\hline 
Number of RF chains & $S=16$\tabularnewline
\hline 
Carrier frequency & $f_{c}=28$ GHz\tabularnewline
\hline 
Pilot length & $T=8$\tabularnewline
\hline 
Under-sampling ratio & $\frac{M}{N}=50\%$\tabularnewline
\hline 
Indices of Active BSs & $3,4,5,6$\tabularnewline
\hline 
DeepMIMO scenario & Outdoor1 (O1) Scenario\tabularnewline
\hline 
Active users & $R1000-R1300$\tabularnewline
\hline 
ULA size & $(N_{x},N_{y},N_{z})=(1,256,1)$\tabularnewline
\hline 
Number of paths & 3\tabularnewline
\hline 
\end{tabular}
\end{table}

\subsection{Validation of Assumptions and Theoretical Analysis\label{verify_Oracle}}

\subsubsection{Assumptions and Oracle Inequality Verifications}

\begin{figure}
\subfloat[\label{verify_Assum2}]{\begin{centering}
\includegraphics[viewport=0bp 0bp 190bp 312bp,clip,scale=0.55]{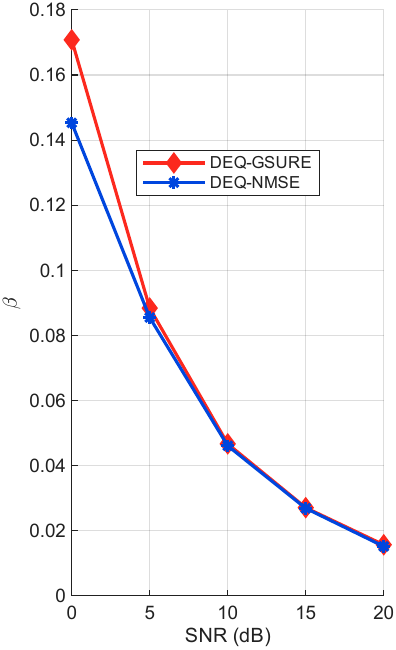}
\par\end{centering}
}\subfloat[\label{verify_Assum3}]{\begin{centering}
\includegraphics[viewport=190bp 0bp 406bp 312bp,clip,scale=0.55]{plot/verify_ass/varify_ass2_3}
\par\end{centering}
}

\caption{(a) $\beta$ (defined in \eqref{beta}) comparison at different SNR
levels; (b) Impact of network bias on the performance and $\omega$
(defined in \eqref{beta}) comparison at different SNR levels.}
\end{figure}

To validate the assumptions used in our theoretical analysis, we conduct
experiments on synthetic channel data with a sparsity level of $k=3$.
In Fig. \ref{verify_Assum2}, we compare the average $\beta$ in Assumption
\ref{good_pmse} for the GSURE- and MSE-optimal solutions at different
SNR levels. It is observed that $\beta\ll1$ is easily satisfied for
both solutions. Although good PMSE performance can be achieved for
the proposed DEQ-GSURE, a gap remains between PMSE and MSE due to
the residual error term in \eqref{MSE}. Therefore, it is necessary
to quantify the MSE discrepancy between the GSURE- and MSE-optimal
solutions. Fig. \ref{verify_Assum3} illustrates the impact of network
bias on the recovery performance of both DEQ-GSURE and DEQ-NMSE, validating
Assumption \ref{zero_output} (i.e., $R_{\bm{\Theta}}(\mathbf{0})=\mathbf{0}$)
and comparing the average $\omega$ (i.e., NMSE values) from Assumption
\ref{good_pmse} at various SNR levels. The results show that $\omega\ll1$
is readily achievable. Moreover, it shows that forcing zero network
bias leads to only subtle performance degradation, confirming that
a bias-free mapping is a practically sound assumption. Furthermore,
DEQ-GSURE exhibits comparable performance to DEQ-NMSE on synthetic
data, and the gap diminishes as SNR increases. This trend further
validates the correctness of the theoretical result in Theorem \ref{oracle inequality}.

\subsubsection{Empirical Validation of $\ell_{1/2}$ Sparsity and SGF Bound}

\begin{figure}[!t]
\begin{centering}
\includegraphics[scale=0.55]{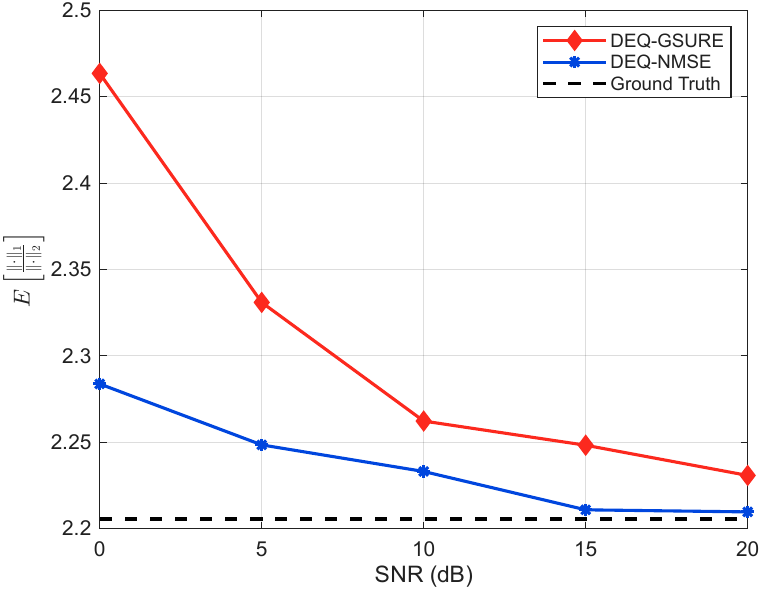}
\par\end{centering}
\centering{}\caption{Empirical $\mathbb{E}\left[\frac{\Vert\cdot\Vert_{1}}{\Vert\cdot\Vert_{2}}\right]$
at different SNR levels.\label{verify_Sparse}}
\end{figure}
\begin{figure}[!t]
\begin{centering}
\includegraphics[scale=0.55]{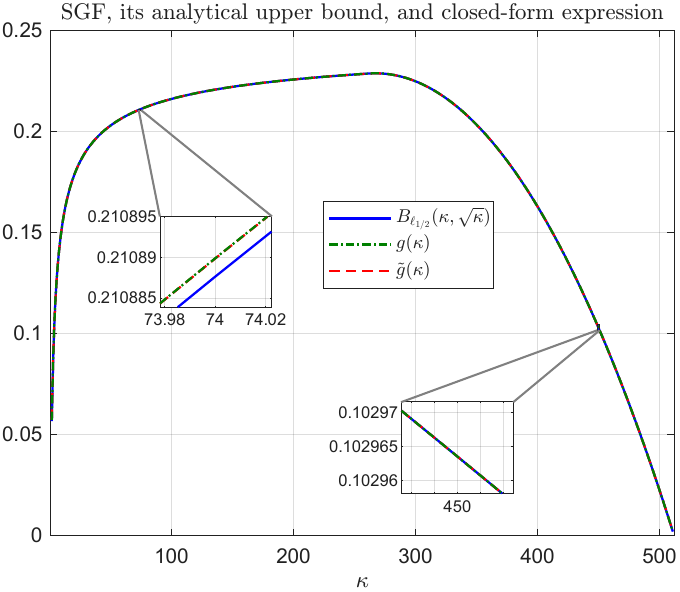}
\par\end{centering}
\centering{}\caption{Empirical verification of SGF, its upper bound, and closed-form expression.
\label{fig: compare 3 function}}
\end{figure}

We validate the $\ell_{1/2}$ sparsity of GSURE- and MSE-optimal DEQ
outputs using synthetic channels with sparsity $k=3$. Fig. \ref{verify_Sparse}
shows the expected sparsity $\mathbb{E}\left[\frac{\Vert\cdot\Vert_{1}}{\Vert\cdot\Vert_{2}}\right]$
under varying SNRs for DEQ-GSURE, DEQ-NMSE, and the ground truth.
It shows that both the GSURE-optimal and MSE-optimal solutions exhibit
strong sparsity and compressibility. This is attributed to the contraction
property of the DEQ network and the effectiveness of the employed
loss functions, i.e., GSURE and MSE, in promoting compressible solutions.
MSE-optimal results align more closely with ground-truth sparsity.
As SNR increases, both outputs converge to the true sparsity, consistent
with Theorem \ref{bounded l_1/2}. In Fig. \ref{fig: compare 3 function},
we plot the SGF function $B_{\ell_{1/2}}(\kappa,\sqrt{\kappa}$),
its bound $g(\kappa$) from \eqref{eq: gk}, and a closed-form $\tilde{g}(\kappa)$
(see \eqref{eq:closed form gk}). The bound is nearly tight across
$\kappa$, especially for $N\ll\kappa\le2N$; for $0\leq\kappa\ll N,$
the gap grows slightly but stays small. $\tilde{g}(\kappa$) effectively
tracks $g(\kappa)$, validating its use in bounding residual energy
in practice.

\subsubsection{Unbiasedness of GSURE}

\begin{figure}
\begin{centering}
\includegraphics[clip,scale=0.35]{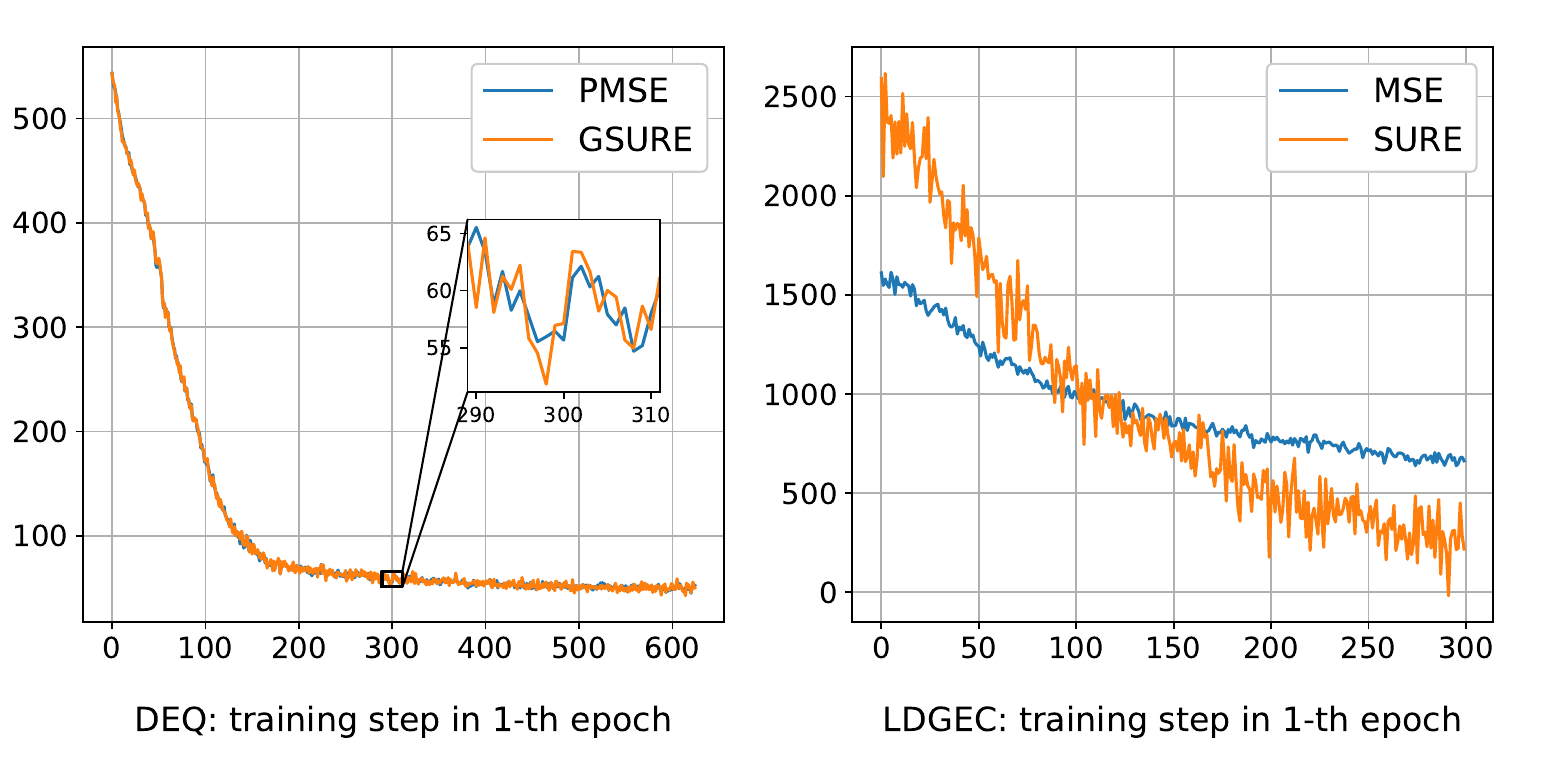}
\par\end{centering}
\centering{}\caption{Training curves of DEQ-PMSE, DEQ-GSURE, LDGEC-NMSE, and LDGEC-SURE.
\label{Accuracy}}
\end{figure}

Although GSURE and SURE are unbiased estimates of PMSE and MSE, respectively,
they do not capture second-order statistics, potentially causing discrepancies
during training. In particular, using SURE as a surrogate for MSE
requires converting the original linear model \eqref{CS} into an
AWGN model, which demands accurate noise variance estimation, and
is extremely challenging in practice. In contrast, GSURE operates
directly on the original model, offering closer alignment with PMSE
during training. As illustrated in Fig. \ref{Accuracy}, DEQ trained
with supervised PMSE and unsupervised GSURE exhibit nearly identical
curves, while LDGEC models show a clear mismatch between supervised
MSE and unsupervised SURE. This highlights GSURE\textquoteright s
robustness as a training loss. However, a fundamental gap remains
between PMSE and MSE, necessitating a quantification of the performance
difference between GSURE-trained and oracle MSE-trained DEQ.

\subsubsection{Types of Measurement Matrix}

\begin{figure}
\begin{centering}
\includegraphics[scale=0.55]{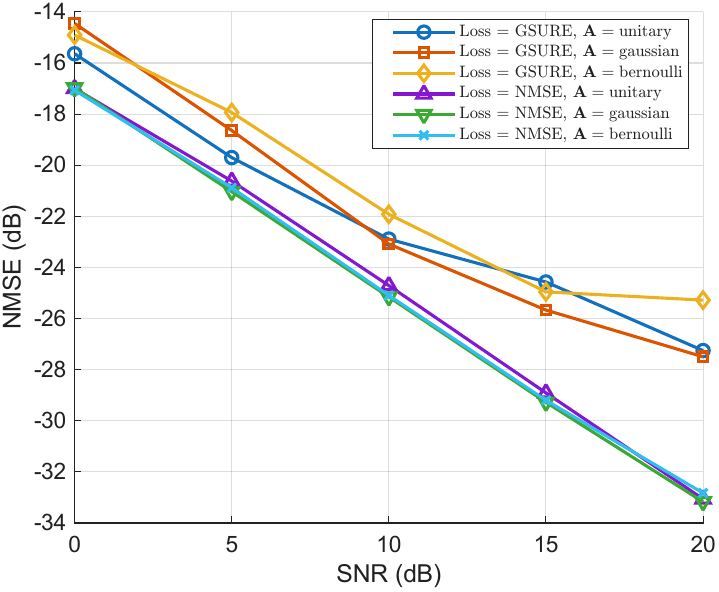}
\par\end{centering}
\caption{NMSE performance comparison for different types of measurement matrices.\label{fig:diff_A}}
\end{figure}

The channel recovery performance of DEQ-NMSE and DEQ-GSURE under different
types of measurement matrices is shown in Fig. \ref{fig:diff_A}.
It demonstrates that both learning methods are insensitive to the
choice of measurement matrix, highlighting the strong representation
capacity and adaptability of the DEQ architecture. Given that the
partial orthogonal matrix simplifies loss function design and facilitates
theoretical analysis without significant loss of performance, we adopt
it throughout this paper.

\subsection{LCE Performance Evaluation}

\subsubsection{Channel Recovery Accuracy}

\begin{figure}[!t]
\begin{centering}
\includegraphics[scale=0.42]{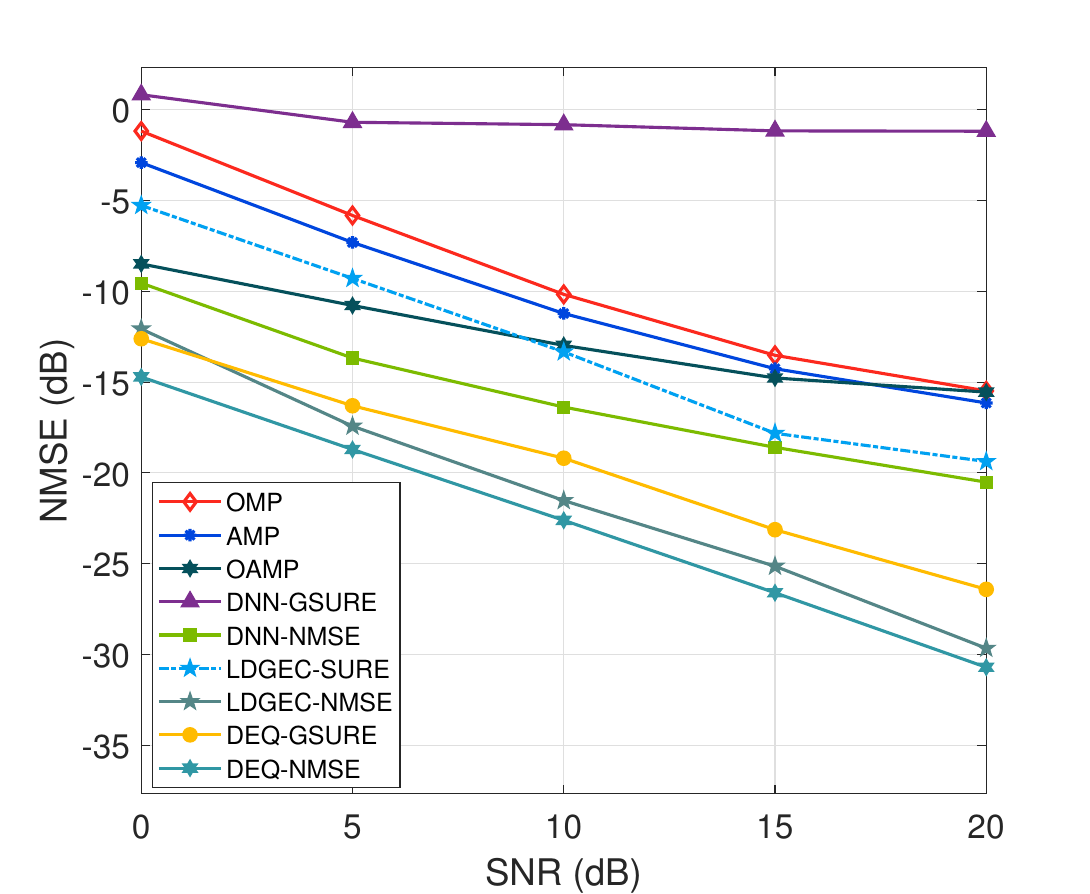}
\par\end{centering}
\centering{}\caption{NMSE performance comparison at different SNR levels.\label{NMSE_performance}}
\end{figure}

In Fig. \ref{NMSE_performance}, we show the NMSE performance of various
methods across different SNR levels. First, the proposed DEQ-GSURE
significantly outperforms classical algorithms such as OMP, AMP, and
OAMP. Second, it surpasses DNN-based methods, including both unsupervised
and supervised versions, thanks to the fixed-point network's convergence-aware
design and the GSURE loss, which offers performance guarantees for
UCL. Third, compared with the model-driven LDGEC-SURE, the proposed
DEQ-GSURE achieves better UL performance (e.g., an 8 dB gain at 0
dB SNR), mainly because LDGEC uses a DU architecture whose performance
is limited by the number of iterations, while the DEQ architecture
is iteration-free, and LDGEC employs a SURE loss restricted to the
AWGN model, which may not hold during training. Even in SCL mode,
DEQ-NMSE maintains a clear advantage over LDGEC-NMSE. Finally, DEQ-GSURE
achieves performance close to its supervised counterpart, confirming
the effectiveness of the proposed UCL approach.

\subsubsection{Out-of-scenario Performance}

\begin{figure}[!t]
\begin{centering}
\includegraphics[scale=0.42]{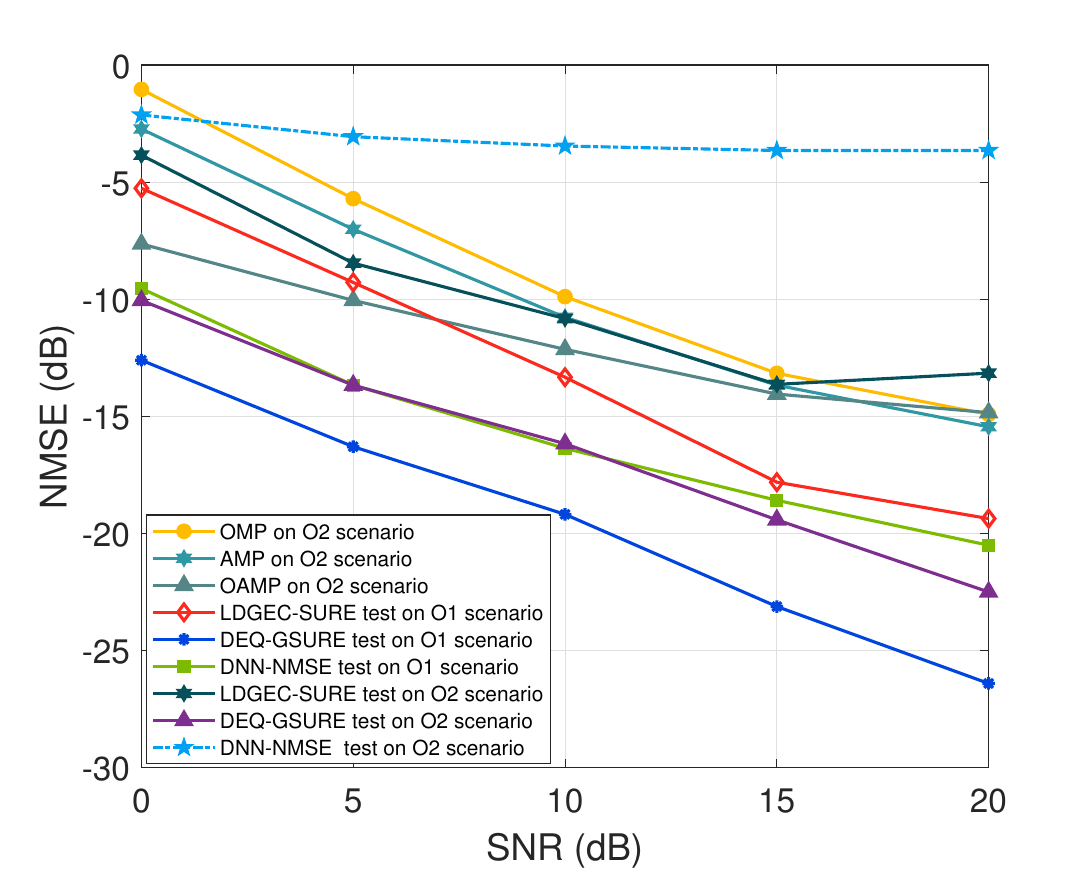}
\par\end{centering}
\centering{}\caption{Out-of-scenario performance of DEQ-GSURE and LDGEC-SURE.\label{robust}}
\end{figure}

We consider out-of-distribution data, where training data corresponds
to the Outdoor1 (O1) scenario in DeepMIMO, while the test data is
selected from other scenarios, i.e., the Outdoor2 (O2) Dynamic Scenario.
Fig. \ref{robust} shows that the performance of all algorithms degrades
when there are data mismatches. However, the proposed DEQ-GSURE exhibits
stronger robustness and generalizability to out-of-scenario data compared
to baselines. Consequently, the proposed UCL scheme can be effectively
deployed in practical communication systems with dynamic environments.

\section{Conclusions}

\label{conclusion} In this paper, we proposed GSURE-based Unsupervised
Deep Equilibrium Learning (GUDL) for large-scale MIMO channel estimation
without requiring ground-truth labels. By combining GSURE as a surrogate
loss and the contractive DEQ architecture, the method achieves near-oracle
performance. A theoretical analysis based on an oracle inequality
reveals that the performance gap to supervised learning can vanish
with the RIP constant approaching zero. Extensive experiments not
only demonstrate the effectiveness and robustness of GUDL, but also
verify the theoretical predictions, showing that our DEQ performs
comparably to its supervised counterparts across various SNR levels.

\appendices{}

\section{Proof of Theorem \ref{SGF Upper Bound}\label{proof of SGF Upper Bound}}

The process of finding a bound of $B_{\ell_{1/2}}(\kappa,\sqrt{\kappa})$
is equivalent to solving 
\begin{align}
\max_{\bm{\alpha}\in\mathbb{R}^{2N}} & ~\ \frac{\Vert\bm{\alpha}_{\kappa^{c}}\Vert_{2}^{2}}{\Vert\bm{\alpha}\Vert_{2}^{2}}\label{origin_problem}\\
\text{s.t.} & ~\ \frac{\Vert\bm{\alpha}\Vert_{1}}{\Vert\bm{\alpha}\Vert_{2}}\le\sqrt{\kappa},\nonumber \\
 & ~\ \Vert\bm{\alpha}_{\kappa^{c}}\Vert_{\infty}\le\frac{\Vert\bm{\alpha}_{\kappa}\Vert_{1}}{\kappa}.\nonumber 
\end{align}
Such optimization (\ref{origin_problem}) is hard to solve and to
derive a closed form since it is non-convex and suffers from too many
optimization variables, i.e., $2N$ variables. However, the intuition
is that those $\bm{\alpha}\in\mathbb{R}^{2N}$ that achieve $\max\Vert\bm{\alpha}_{\kappa^{c}}\Vert_{2}^{2}/\Vert\bm{\alpha}\Vert_{2}^{2}$
are some special points, which motivates us to constrain the structure
of $\bm{\alpha}$ step by step to reduce optimization variables and
simplify the optimization problem.

Therefore, our proof proceeds as follows:
\begin{enumerate}
\item We show that (\ref{origin_problem}) can be reduced to 
\begin{align}
\max_{\tilde{\bm{\alpha}}} & ~\ \frac{\Vert\tilde{\bm{\alpha}}_{\kappa^{c}}\Vert_{2}^{2}}{\Vert\tilde{\bm{\alpha}}\Vert_{2}^{2}},\label{sorted_decreasing}\\
\text{s.t.} & ~\ \frac{\Vert\tilde{\bm{\alpha}}\Vert_{1}}{\Vert\tilde{\bm{\alpha}}\Vert_{2}}\le\sqrt{\kappa},\nonumber 
\end{align}
where 
\begin{equation}
\tilde{\bm{\alpha}}=\begin{bmatrix}\tilde{\bm{\alpha}}_{\kappa}\\
\tilde{\bm{\alpha}}_{\kappa^{c}}
\end{bmatrix}=\begin{bmatrix}\begin{bmatrix}\alpha(1),\alpha(2),\dots,\alpha(\kappa)\end{bmatrix}^{T}\\
\begin{bmatrix}\alpha(\kappa+1),\alpha(\kappa+2),\dots,\alpha(2N)\end{bmatrix}^{T}
\end{bmatrix}
\end{equation}
and $\alpha(\cdot)\in\mathcal{A}$ satisfies $\alpha(i)\ge0$, $\alpha(i+1)-\alpha(i)\le0$
for any $i\in{1,2,\dots,2N}$. Such optimization \eqref{sorted_decreasing}
ensures that we reduce the search space from $\mathbb{R}^{2N}$ to
the space of all non-negative vectors sorted in decreasing order of
magnitude, characterized by $\alpha(\cdot)\in\mathcal{A}$.
\item Any $\tilde{\bm{\alpha}}$ that achieves $\sup_{\alpha\in\mathcal{A}}\Vert\tilde{\bm{\alpha}}_{\kappa^{c}}\Vert_{2}^{2}/\Vert\tilde{\bm{\alpha}}\Vert_{2}^{2}$
has the same structure as $\bm{\alpha}^{\star}$, where $\bm{\alpha}^{\star}$
holds that
\begin{enumerate}
\item \label{alpha_T}$\bm{\alpha}_{\kappa}^{\star}=C\cdot\begin{bmatrix}1,\rho,\rho,\dots,\rho\end{bmatrix}^{T}$,
$C>0$ is a constant;
\item \label{alpha_Tc}$\bm{\alpha}_{\kappa^{c}}^{\star}=C\cdot[\underbrace{\rho,\rho,\rho,\dots,\rho}_{p},\lambda\rho,0,\dots,0]^{T}$,
$C>0$ is a constant, $p$ is a non-negative integer, $0\le\rho,\lambda\le1$
denote the dynamic range, which will be introduced later. 
\end{enumerate}
Based on $\bm{\alpha}^{\star}$, \eqref{sorted_decreasing} can be
reduced to 
\begin{align}
\max_{\bm{\alpha}^{\star}} & ~\ \frac{\Vert\bm{\alpha}_{\kappa^{c}}^{\star}\Vert_{2}^{2}}{\Vert\bm{\alpha}^{\star}\Vert_{2}^{2}}\label{alpha_star}\\
\text{s.t.} & ~\ \frac{\Vert\bm{\alpha}^{\star}\Vert_{1}}{\Vert\bm{\alpha}^{\star}\Vert_{2}}\le\sqrt{\kappa}.\nonumber 
\end{align}

\item Since any $\bm{\alpha}^{\star}$ can be characterized by 3 parameters,
i.e., $(\rho,p,\lambda)$, \eqref{alpha_star} is equivalent to 
\begin{align}
\max_{p,\rho,\lambda} & ~\ \frac{(p+\lambda^{2})\cdot\rho^{2}}{1+(p+\lambda^{2}+\kappa-1)\cdot\rho^{2}}\label{rho_p_lambda}\\
\text{s.t.} & ~\ \frac{1+(p+\kappa+\lambda-1)\cdot\rho}{\sqrt{1+(p+\kappa+\lambda^{2}-1)\cdot\rho^{2}}}\le\sqrt{\kappa}\nonumber \\
 & ~\ 0\le p\le2N-\kappa-1,\quad p\in\mathbb{N}^{+}\nonumber \\
 & ~\ 0\le\lambda,\rho\le1.\nonumber 
\end{align}
Relax the first constraint in \eqref{rho_p_lambda} with $p+\lambda^{2}\le p+\lambda$
and define $x=p+\lambda^{2}$, then $x\in\mathbb{R}^{+}$ and $x\in(0,2N-\kappa)$,
allowing the elimination of the integer-optimized variable $p$. The
final optimization can be reduced to 
\begin{equation}
\max_{x\in(0,2N-\kappa)}\frac{x\cdot\tilde{\rho}^{2}(x)}{1+(x+\kappa-1)\cdot\tilde{\rho}^{2}(x)},\label{obj_relax}
\end{equation}
where 
\[
\tilde{\rho}(x)=\frac{\sqrt{\frac{x\cdot\kappa}{x+\kappa-1}}-1}{x-1}
\]
is derived from the first constraint in \eqref{rho_p_lambda}. We
show that \eqref{obj_relax} has a unique maximum and the relaxation
causes negligible difference. \\
\end{enumerate}
Proofs:
\begin{enumerate}
\item Since any other $\bm{\alpha}\in\mathbb{R}^{2N}$ can be converted
to $\tilde{\bm{\alpha}}$ by first inverting the negative values to
positive and then sorting the elements by their values, the following
equation still holds: 
\begin{equation}
\frac{\Vert\tilde{\bm{\alpha}}\Vert_{1}}{\Vert\tilde{\bm{\alpha}}\Vert_{2}}=\frac{\Vert\bm{\alpha}\Vert_{1}}{\Vert\bm{\alpha}\Vert_{2}},\quad\frac{\Vert\tilde{\bm{\alpha}}_{\kappa^{c}}\Vert_{2}}{\Vert\tilde{\bm{\alpha}}\Vert_{2}}=\frac{\Vert\bm{\alpha}_{\kappa^{c}}\Vert_{2}}{\Vert\bm{\alpha}\Vert_{2}}.
\end{equation}
\item Proof of \ref{alpha_T}): We prove it by contradiction. The insight
is that assuming another $\bm{\alpha}'=\begin{bmatrix}{\bm{\alpha}'_{\kappa}}^{T},{\bm{\alpha}'_{\kappa^{c}}}^{T}\end{bmatrix}^{T}\neq\bm{\alpha}^{\star}$
reaches maximum $\Vert\bm{\alpha}'_{\kappa^{c}}\Vert_{2}/\Vert\bm{\alpha}'_{\kappa}\Vert_{2}$
with any $\alpha'\in\mathcal{A}$, if we could construct a new vector
$\bm{\alpha}'(\varsigma)=\begin{bmatrix}{\bm{\alpha}_{\kappa}(\varsigma)}^{T},{\bm{\alpha}'_{\kappa^{c}}}^{T}\end{bmatrix}^{T}$
from $\bm{\alpha}'$ through some operation and prove 
\begin{equation}
\begin{cases}
{\displaystyle \frac{\Vert\bm{\alpha}'(\varsigma)\Vert_{1}}{\Vert\bm{\alpha}'(\varsigma)\Vert_{2}}\le\frac{\Vert\bm{\alpha}'\Vert_{1}}{\Vert\bm{\alpha}'\Vert_{2}},}\\[10pt]
{\displaystyle \frac{\Vert\bm{\alpha}'_{\kappa^{c}}\Vert_{2}}{\Vert\bm{\alpha}_{\kappa}(\varsigma)\Vert_{2}}\ge\frac{\Vert\bm{\alpha}'_{\kappa^{c}}\Vert_{2}}{\Vert\bm{\alpha}'_{\kappa}\Vert_{2}},}\\[10pt]
\bm{\alpha}'(\varsigma)=\bm{\alpha}',~\ {\text{if and only if}}\ \bm{\alpha}'=\bm{\alpha}^{\star},
\end{cases}
\end{equation}
then the proof could be accomplished. Now we aim at constructing such
$\varsigma$. Let 
\begin{equation}
\bm{\alpha}'_{\kappa}=\begin{bmatrix}\alpha'(1),\alpha'(2),\dots,\alpha'(\kappa)\end{bmatrix}^{T}
\end{equation}
and introduce $\bm{\alpha}_{\kappa}(\varsigma)$, 
\begin{align}
\bm{\alpha}_{\kappa}(\varsigma)= & \begin{bmatrix}\alpha'(1)+\varsigma,\alpha'(\kappa),\dots,\alpha'(\kappa)\end{bmatrix}^{T}\nonumber \\
= & \bm{\alpha}_{\kappa}^{\star}+\varsigma\cdot\begin{bmatrix}1,0,\dots,0\end{bmatrix}^{T},\quad\varsigma>0,
\end{align}
where $C=\alpha'(1)$ in the above $\bm{\alpha}_{\kappa}^{\star}$.
Let $\Vert\bm{\alpha}_{\kappa}(\varsigma)\Vert_{2}\le\Vert\bm{\alpha}'_{\kappa}\Vert_{2}$,
\[
\hspace{-4.8em}\Longrightarrow\Vert\bm{\alpha}_{\kappa}^{\star}\Vert_{2}^{2}+\varsigma^{2}+2\alpha'(1)\cdot\varsigma\le\Vert\bm{\alpha}'_{\kappa}\Vert_{2}^{2}
\]
\begin{equation}
\Longrightarrow\varsigma\le\sqrt{[\alpha'(1)]^{2}+\Vert\bm{\alpha}'_{\kappa}\Vert_{2}^{2}-\Vert\bm{\alpha}_{\kappa}^{\star}\Vert_{2}^{2}}-\alpha'(1)=\varsigma^{\star}.
\end{equation}
If we denote 
\begin{align}
 & \bm{\alpha}'_{\kappa}=\bm{\alpha}_{\kappa}^{\star}+\bm{\delta}=\begin{bmatrix}\alpha'(1),\alpha'(\kappa),\dots,\alpha'(\kappa)\end{bmatrix}^{T}\nonumber \\
 & +\begin{bmatrix}0,\alpha'(2)-\alpha'(\kappa),\dots,\alpha'(\kappa-1)-\alpha'(\kappa),0\end{bmatrix}^{T},
\end{align}
then we can rewrite 
\begin{align}
\varsigma^{\star}= & \sqrt{[\alpha'(1)]^{2}+\Vert\bm{\delta}\Vert_{2}^{2}+2{\bm{\alpha}_{\kappa}^{\star}}^{T}\bm{\delta}}-\alpha'(1)\nonumber \\
= & \sqrt{[\alpha'(1)]^{2}+\Vert\bm{\delta}\Vert_{2}^{2}+2\alpha'(\kappa)\cdot\Vert\bm{\delta}\Vert_{1}}-\alpha'(1).\label{varsigma}
\end{align}
Using $\Vert\bm{\delta}\Vert_{2}\le\Vert\bm{\delta}\Vert_{1}$ and
$\alpha'(\kappa)\le\alpha'(1)$, we can derive 
\begin{align}
 & \Vert\bm{\alpha}_{\kappa}(\varsigma)\Vert_{1}\nonumber \\
\le & \Vert\bm{\alpha}_{\kappa}^{\star}\Vert_{1}+\varsigma^{\star}\nonumber \\
\le & \Vert\bm{\alpha}_{\kappa}^{\star}\Vert_{1}+\sqrt{[\alpha'(1)]^{2}+\Vert\bm{\delta}\Vert_{1}^{2}+2\alpha'(1)\cdot\Vert\bm{\delta}\Vert_{1}}-\alpha'(1)\nonumber \\
= & \Vert\bm{\alpha}_{\kappa}^{\star}\Vert_{1}+\sqrt{\left(\alpha'(1)+\Vert\bm{\delta}\Vert_{1}\right)^{2}}-\alpha'(1)\nonumber \\
= & \Vert\bm{\alpha}_{\kappa}^{\star}\Vert_{1}+\Vert\bm{\delta}\Vert_{1}=\Vert\bm{\alpha}'_{\kappa}\Vert_{1}.\label{exact_less}
\end{align}
Equality only holds if $\bm{\delta}=\bm{0}$, otherwise $\Vert\bm{\alpha}_{\kappa}(\varsigma)\Vert_{1}<\Vert\bm{\alpha}'_{\kappa}\Vert_{1}$
with an exact non-negative gap. However, $\varsigma=0$ if and only
if $\bm{\delta}=\bm{0}$, in which case $\bm{\alpha}_{\kappa}(\varsigma)=\bm{\alpha}'_{\kappa}$.
For the case $\bm{\delta}\neq\bm{0}$, we introduce a differentiable
function $\ell(\varsigma)$ defined as 
\begin{align}
\ell(\varsigma)= & \frac{\Vert\bm{\alpha}'(\varsigma)\Vert_{1}}{\Vert\bm{\alpha}'(\varsigma)\Vert_{2}}\nonumber \\
= & \frac{\Vert\bm{\alpha}_{\kappa}^{\star}\Vert_{1}+\varsigma+\Vert\bm{\alpha}'_{\kappa^{c}}\Vert_{1}}{\sqrt{\Vert\bm{\alpha}_{\kappa}^{\star}\Vert_{2}^{2}+\varsigma^{2}+2\alpha'(1)\varsigma+\Vert\bm{\alpha}'_{\kappa}\Vert_{2}^{2}}}.
\end{align}
Recalling \eqref{exact_less}, we have 
\begin{equation}
\ell(\varsigma^{\star})=\frac{\Vert\bm{\alpha}'(\varsigma^{\star})\Vert_{1}}{\Vert\bm{\alpha}'(\varsigma^{\star})\Vert_{2}}=\frac{\Vert\bm{\alpha}_{\kappa}^{\star}\Vert_{1}+\varsigma^{\star}+\Vert\bm{\alpha}'_{\kappa^{c}}\Vert_{1}}{\sqrt{\Vert\bm{\alpha}'_{\kappa}\Vert_{2}^{2}+\Vert\bm{\alpha}'_{\kappa^{c}}\Vert_{2}^{2}}}<\frac{\Vert\bm{\alpha}'\Vert_{1}}{\Vert\bm{\alpha}'\Vert_{2}},
\end{equation}
where the gap $\Vert\bm{\alpha}'\Vert_{1}/\Vert\bm{\alpha}'\Vert_{2}-\ell(\varsigma^{\star})$
is \textbf{strictly positive}. Therefore, we can always find a $\Delta\in(0,\Delta_{\max})$
such that 
\begin{equation}
\ell(\varsigma^{\star}-\Delta)\le\frac{\Vert\bm{\alpha}'\Vert_{1}}{\Vert\bm{\alpha}'\Vert_{2}}.
\end{equation}
Using $\Vert\bm{\alpha}'(\varsigma^{\star})\Vert_{2}=\Vert\bm{\alpha}'_{\kappa}\Vert_{2}$,
the above $\Delta_{\max}$ can be solved by 
\begin{equation}
\ell(\varsigma^{\star}-\Delta)=\frac{\Vert\bm{\alpha}'\Vert_{1}}{\Vert\bm{\alpha}'\Vert_{2}},
\end{equation}
\begin{equation}
\Longrightarrow\frac{\Vert\bm{\alpha}'(\varsigma^{\star})\Vert_{1}-\Delta}{\sqrt{\Vert\bm{\alpha}'\Vert_{2}^{2}+\Delta^{2}-2\Delta\left(\varsigma^{\star}+\alpha'(1)\right)}}=\frac{\Vert\bm{\alpha}'\Vert_{1}}{\Vert\bm{\alpha}'\Vert_{2}},
\end{equation}
\begin{align}
\Longrightarrow & \left(\frac{\Vert\bm{\alpha}'\Vert_{1}^{2}}{\Vert\bm{\alpha}'\Vert_{2}^{2}}-1\right)\Delta^{2}+\nonumber \\
 & 2\Delta\left(\Vert\bm{\alpha}'(\varsigma^{\star})\Vert_{1}-\frac{\Vert\bm{\alpha}'\Vert_{1}^{2}}{\Vert\bm{\alpha}'\Vert_{2}^{2}}(\varsigma^{\star}+\alpha'(1))\right)+\nonumber \\
 & \Vert\bm{\alpha}'\Vert_{1}^{2}-\Vert\bm{\alpha}'(\varsigma^{\star})\Vert_{1}^{2}=0.\label{delta_solve}
\end{align}
By introducing new notations, the above formula can be shortened as
\begin{equation}
\zeta(\Delta)=\zeta_{a}\Delta^{2}+2\zeta_{b}\Delta+\zeta_{c}=0.
\end{equation}
Note that in \eqref{delta_solve},
\begin{equation}
\begin{cases}
\zeta_{a}=\Vert\bm{\alpha}'\Vert_{1}^{2}/\Vert\bm{\alpha}'\Vert_{2}^{2}-1>0,\\
\zeta_{c}=\Vert\bm{\alpha}'\Vert_{1}^{2}-\Vert\bm{\alpha}'(\varsigma^{\star})\Vert_{1}^{2}>0.
\end{cases}
\end{equation}
According to the properties of quadratic equations, the $\Delta_{\max}$
can be determined by 
\begin{align}
 & \Delta_{\max}=\nonumber \\
 & \begin{cases}
{\displaystyle \varsigma^{\star},} & {\text{if}}\ \zeta_{b}\ge0\\
\varsigma^{\star}, & {\text{if}}\ \zeta_{b}\le0~\ {\text{and}}~\ \zeta({\displaystyle -\frac{\zeta_{b}}{2\zeta_{a}})\ge0}\\
{\displaystyle \frac{-\zeta_{b}+\sqrt{\zeta_{b}^{2}-\zeta_{a}\zeta_{c}}}{\zeta_{a}},} & {\text{if}}\ \zeta_{b}\le0~\ {\text{and}}~\ \zeta({\displaystyle -\frac{\zeta_{b}}{2\zeta_{a}})<0.}
\end{cases}
\end{align}
In summary, for any $\bm{\alpha}'\neq\bm{\alpha}^{\star}$, we always
could find a $\varsigma$ by setting 
\begin{equation}
\varsigma\in\begin{cases}
\{0\}, & {\text{if}}\ \bm{\delta}=\bm{0},\\
(\varsigma^{\star}-\Delta_{\max},\varsigma^{\star}), & {\text{if}}\ \bm{\delta}\neq\bm{0},
\end{cases}
\end{equation}
and construct a new vector $\bm{\alpha}'(\varsigma)$ such that 
\begin{equation}
\begin{cases}
{\displaystyle \ell(\varsigma)\le\frac{\Vert\bm{\alpha}'\Vert_{1}}{\Vert\bm{\alpha}'\Vert_{2}},}\\[8pt]
{\displaystyle \frac{\Vert\bm{\alpha}'_{\kappa^{c}}\Vert_{2}}{\Vert\bm{\alpha}_{\kappa}(\varsigma)\Vert_{2}}\ge\frac{\Vert\bm{\alpha}'_{\kappa^{c}}\Vert_{2}}{\Vert\bm{\alpha}_{\kappa}(\varsigma^{\star})\Vert_{2}}\ge\frac{\Vert\bm{\alpha}'_{\kappa^{c}}\Vert_{2}}{\Vert\bm{\alpha}'_{\kappa}\Vert_{2}},}
\end{cases}
\end{equation}
which contradicts the assumption according to the arbitrariness of
$\alpha(\cdot)\in\mathcal{A}$ and $\bm{\alpha}'_{\kappa}$.\\

Proof of \ref{alpha_Tc}): Based on the proof of \ref{alpha_T}),
we prove \ref{alpha_Tc}) also by contradiction. Let 
\begin{equation}
\bm{\alpha}'_{\kappa^{c}}=\begin{bmatrix}\alpha'(\kappa+1),\alpha'(\kappa+2),\dots,\alpha'(2N)\end{bmatrix}^{T}
\end{equation}
and introduce $\bm{\alpha}_{\kappa^{c}}(p,\lambda)$ 
\begin{equation}
\bm{\alpha}_{\kappa^{c}}(p,\lambda)=\begin{bmatrix}\smash[b]{\underbrace{\rho,\dots,\rho}_{p}},\lambda\rho,0,0,\dots,0\end{bmatrix}^{T},
\end{equation}
\\
 where we assume $\alpha'(1)=1$ without loss of generality and $p,\lambda$
are obtained respectively by 
\begin{equation}
p=\Big\lfloor\frac{\Vert\bm{\alpha}'_{\kappa^{c}}\Vert_{1}}{\rho}\Big\rfloor,\quad\lambda=\frac{\Vert\bm{\alpha}'_{\kappa^{c}}\Vert_{1}-p\rho}{\rho}.\label{p_lamb}
\end{equation}
According to Theorem 4.1 in \cite{hurley2009comparing}, which characterizes
a property of the metric defined by the ratio of $\ell_{1}$ and $\ell_{2}$
norms, the following holds: 
\begin{equation}
\Vert\bm{\alpha}_{\kappa^{c}}(p,\lambda)\Vert_{2}\ge\Vert\bm{\alpha}'_{\kappa}\Vert_{2}.\label{L2_increase}
\end{equation}
Based on \eqref{p_lamb}, we have 
\begin{equation}
\Vert\bm{\alpha}_{\kappa^{c}}(p,\lambda)\Vert_{1}=p\cdot\rho+\lambda\rho=\Vert\bm{\alpha}'_{\kappa^{c}}\Vert_{1}.\label{L1_constant}
\end{equation}
\eqref{L2_increase}-\eqref{L1_constant} imply that given any $\bm{\alpha}'_{\kappa^{c}}$,
we can construct $\bm{\alpha}'(p,\lambda)=\begin{bmatrix}{\bm{\alpha}'_{\kappa}}^{T},{\bm{\alpha}_{\kappa^{c}}(p,\lambda)}^{T}\end{bmatrix}^{T}$
and conclude 
\begin{equation}
\begin{cases}
{\displaystyle \frac{\Vert\bm{\alpha}'(p,\lambda)\Vert_{1}}{\Vert\bm{\alpha}'(p,\lambda)\Vert_{2}}=\frac{\Vert\bm{\alpha}'\Vert_{1}}{\sqrt{\Vert\bm{\alpha}'_{\kappa}\Vert_{2}^{2}+\Vert\bm{\alpha}_{\kappa^{c}}(p,\lambda)\Vert_{2}^{2}}}\le\frac{\Vert\bm{\alpha}'\Vert_{1}}{\Vert\bm{\alpha}'\Vert_{2}},}\\[10pt]
{\displaystyle \frac{\Vert\bm{\alpha}_{\kappa^{c}}(p,\lambda)\Vert_{2}}{\Vert\bm{\alpha}'_{\kappa}\Vert_{2}}\ge\frac{\Vert\bm{\alpha}'_{\kappa^{c}}\Vert_{2}}{\Vert\bm{\alpha}'_{\kappa}\Vert_{2}},}
\end{cases}
\end{equation}
which contradicts the assumption according to the arbitrariness of
$\alpha(\cdot)\in\mathcal{A}$ and $\bm{\alpha}'_{\kappa}$.
\item Now our objective becomes \eqref{alpha_star}, that is
\begin{align}
\max_{p,\rho,\lambda} & ~\ \frac{(p+\lambda^{2})\cdot\rho^{2}}{1+(p+\lambda^{2}+\kappa-1)\cdot\rho^{2}}\label{rho_p_lambda_1}\\
\text{s.t.} & ~\ \frac{1+(p+\kappa+\lambda-1)\cdot\rho}{\sqrt{1+(p+\kappa+\lambda^{2}-1)\cdot\rho^{2}}}\le\sqrt{\kappa}\nonumber \\
 & ~\ 0\le p\le2N-\kappa-1,\quad p\in\mathbb{N}^{+}\nonumber \\
 & ~\ 0\le\lambda,\rho\le1.\nonumber 
\end{align}
We define $x=p+\lambda^{2}$, then $x\in\mathbb{R}$ and $x\in(0,2N-\kappa)$.
Using this substitute and $p+\lambda^{2}\le p+\lambda$ to relax the
first constraint, the objective becomes 
\begin{align}
\max_{x,\rho} & ~\ \frac{x\cdot\rho^{2}}{1+(x+\kappa-1)\cdot\rho^{2}}\\
\text{s.t.} & ~\ \frac{1+(x+\kappa-1)\cdot\rho}{\sqrt{1+(x+\kappa-1)\cdot\rho^{2}}}\le\sqrt{\kappa}\nonumber \\
 & ~\ 0\le x\le2N-\kappa,\quad x\in\mathbb{R}^{+}\nonumber \\
 & ~\ 0\le\rho\le1.\nonumber 
\end{align}
Analyzing this objective function, it is monotonically increasing
with respect to $\rho$ and $x$. However, if $x$ increases, the
upper bound of $\rho$ will decrease. There is a trade-off, and there
exists a best $(x,\rho)$ that maximizes the objective function. According
to the first constraint, we can get: 
\begin{equation}
\frac{1+(x+\kappa-1)\cdot\rho}{\sqrt{1+(x+\kappa-1)\cdot\rho^{2}}}\le\sqrt{\kappa}\Longrightarrow
\end{equation}
\begin{equation}
0\le\rho\le\frac{\sqrt{\frac{x\cdot\kappa}{x+\kappa-1}}-1}{x-1}=\tilde{\rho}(x),
\end{equation}
our goal is to find 
\begin{equation}
\max_{x\in(0,2N-\kappa)}\frac{x\cdot\tilde{\rho}^{2}(x)}{1+(x+\kappa-1)\cdot\tilde{\rho}^{2}(x)}=\max_{x}c(\kappa,x).\label{obj_relax_1}
\end{equation}
In order to solve \eqref{obj_relax_1} through a closed-form solution,
we first take advantage of $t=\sqrt{\frac{x}{x+\kappa-1}}$ to transform
$c(\kappa,x)$ into $c(\kappa,t)$. Let $d=(x+\kappa-1)\cdot\tilde{\rho}^{2}(x)$
:
\begin{align}
d= & (x+\kappa-1)\cdot\left({\displaystyle \frac{\sqrt{\frac{x\cdot\kappa}{x+\kappa-1}}-1}{x-1}}\right)^{2}\nonumber \\
= & \left({\displaystyle \frac{1}{x-1}}\cdot{\displaystyle \frac{x\cdot\kappa-(x+\kappa-1)}{{\displaystyle \sqrt{x\cdot\kappa}}+\sqrt{x+\kappa-1}}}\right)^{2}\nonumber \\
= & \left({\displaystyle \frac{\kappa-1}{{\displaystyle \sqrt{x\cdot\kappa}}+\sqrt{x+\kappa-1}}}\right)^{2}.\label{eq: d_unhandled}
\end{align}
Replacing $x$ with \textbf{$\frac{(\kappa-1)\cdot t^{2}}{1-t^{2}}$},
\eqref{eq: d_unhandled} is reduced to:
\begin{align}
d= & {\displaystyle \frac{(\kappa-1)^{2}}{\left({\displaystyle \sqrt{\frac{(\kappa-1)\cdot t^{2}}{1-t^{2}}\cdot\kappa}}+\sqrt{{\displaystyle \frac{(\kappa-1)\cdot t^{2}}{1-t^{2}}}+\kappa-1}\right)^{2}}}\nonumber \\
= & {\displaystyle \frac{(\kappa-1)(1-t^{2})}{\left(t\sqrt{\kappa}+1\right)^{2}}}.
\end{align}
Therefore, $c(\kappa,x)$ can be converted into $c(\kappa,t)$ through
\begin{align}
c(\kappa,x)= & t^{2}\cdot{\displaystyle \frac{d}{1+d}}\nonumber \\
= & t^{2}\cdot{\displaystyle \frac{\left(\kappa-1\right)\left(1-t^{2}\right)}{\left(t\sqrt{\kappa}+1\right)^{2}+(\kappa-1)\left(1-t^{2}\right)}}\nonumber \\
= & \left(\kappa-1\right)\cdot{\displaystyle \frac{t^{2}\cdot\left(1-t^{2}\right)}{\left(t+\sqrt{\kappa}\right)^{2}}}=c(\kappa,t).\label{c(k,t)}
\end{align}
Since $0\le x\le2N-\kappa$, the value of $t$ lies in $(0,\sqrt{\frac{2N-\kappa}{2N-1}})$.
For any fixed $\kappa$, \eqref{c(k,t)} has a unique stationary point.
Fig. \ref{fig:The-curve-of-c} empirically verifies the uniqueness,
where the red points represent the maximum value points. 
\begin{figure}
\begin{centering}
\includegraphics[scale=0.55]{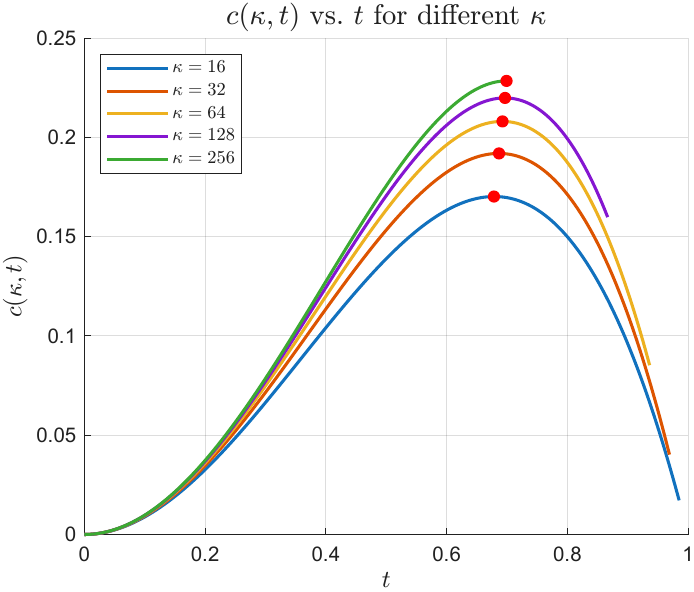}
\par\end{centering}
\caption{The curve of $c(\kappa,t)$ for different $\kappa$.\label{fig:The-curve-of-c}}
\end{figure}
Finally, by setting the derivative to $0$, we have
\begin{equation}
\frac{\partial\ln c(\kappa,t)}{\partial t}=0\Longrightarrow\left({\displaystyle \frac{1}{t}}\right)^{3}-2\left({\displaystyle \frac{1}{t}}\right){\displaystyle +\frac{1}{\sqrt{\kappa}}}=0.\label{depressed_cubic}
\end{equation}
The above cubic equation is said to be \emph{depressed} as the quadratic
term is absent. The discriminant $\left(\frac{-2}{2}\right)^{2}+\left(\frac{1}{3\sqrt{\kappa}}\right)^{3}>0$
indicates that \eqref{depressed_cubic} has three distinct real roots,
among which only one satisfies the constraint on the range of $t$.
In this case, we can adopt the \emph{Trigonometric} solution:
\begin{equation}
t_{k}=\frac{\sqrt{6}}{4\cos\left[\frac{1}{3}\arccos\left({\displaystyle \frac{3\sqrt{6}}{8\sqrt{\kappa}}}\right)-{\displaystyle \frac{2\pi k}{3}}\right]}\quad\text{for}~k=0,1,2.
\end{equation}
Specifically, only $k=0$ meets the requirement. In this way, the
upper bound of the SGF can be expressed in closed form, i.e.,
\begin{equation}
\begin{array}{rl}
B_{\ell_{1/2}}(\kappa,\sqrt{\kappa}) & \le\sup c(\kappa,x)\\
 & ={\displaystyle \frac{(\kappa-1)t^{2}(\kappa)\left(1-t^{2}(\kappa)\right)}{\left(\sqrt{\kappa}+t(\kappa)\right)^{2}}=\tilde{g}(\kappa)},
\end{array}\label{eq:closed form gk}
\end{equation}
with 
\begin{equation}
t(\kappa)=\min\biggr\{\frac{\sqrt{6}}{4\cos\left[\frac{1}{3}\arccos\left({\displaystyle \frac{3\sqrt{6}}{8\sqrt{\kappa}}}\right)\right]},\sqrt{\frac{2N-\kappa}{2N-1}}\biggr\}.
\end{equation}
\end{enumerate}

\section{Proof of Theorem \ref{bounded l_1/2}}

\label{proof_bounded l_1/2}

Let $\mathbf{P}$ be the orthogonal projection matrix associated with
$\mathbf{A},$ i.e., $\mathbf{P}=\mathbf{A}^{T}\left(\mathbf{A}\mathbf{A}^{T}\right)^{-1}\mathbf{A}=\mathbf{A}^{T}\mathbf{A}$.
Then we have that
\begin{align}
\text{\ensuremath{\ell_{1/2}}(\ensuremath{\mathbf{h}_{\mathrm{G}}^{\star}})=}\frac{\Vert\mathbf{h}_{\mathrm{G}}^{\star}\Vert_{1}}{\Vert\mathbf{h}_{\mathrm{G}}^{\star}\Vert_{2}}= & \frac{\Vert\mathbf{P}\mathbf{h}_{\mathrm{G}}^{\star}+(\mathbf{I}_{2N}-\mathbf{P})\mathbf{h}_{\mathrm{G}}^{\star}\Vert_{1}}{\Vert\mathbf{h}^{\star}\Vert_{2}}\nonumber \\
\le & \frac{\Vert\mathbf{P}\mathbf{h}_{\mathrm{G}}^{\star}\Vert_{1}}{\Vert\mathbf{P}\mathbf{h}_{\mathrm{G}}^{\star}\Vert_{2}}+\sqrt{2N}\frac{\Vert(\mathbf{I}_{2N}-\mathbf{P})\mathbf{h}_{\mathrm{G}}^{\star}\Vert_{2}}{\Vert\mathbf{h}^{\star}\Vert_{2}}\nonumber \\
= & \frac{\Vert\mathbf{P}\mathbf{h}_{\mathrm{G}}^{\star}\Vert_{1}}{\Vert\mathbf{P}\mathbf{h}_{\mathrm{G}}^{\star}\Vert_{2}}+\sqrt{2N}\sqrt{1-\frac{\Vert\mathbf{P}\mathbf{h}_{\mathrm{G}}^{\star}\Vert_{2}^{2}}{\Vert\mathbf{h}_{\mathrm{G}}^{\star}\Vert_{2}^{2}}}.\label{unhandled_1}
\end{align}
Recalling Assumption \ref{good_pmse}, we have 
\begin{align}
\Vert\mathbf{P}\mathbf{h}_{\mathrm{G}}^{\star}\Vert_{2}= & \Vert\mathbf{Ph}+\mathbf{P}\mathbf{h}_{\mathrm{G}}^{\star}-\mathbf{Ph}\Vert_{2}\nonumber \\
\ge & \big\vert\Vert\mathbf{Ph}\Vert_{2}-\Vert\mathbf{P}\mathbf{h}_{\mathrm{G}}^{\star}-\mathbf{Ph}\Vert_{2}\big\vert\ge(1-\beta)\Vert\mathbf{Ph}\Vert_{2}.
\end{align}
According to Assumption \ref{zero_output} and the fixed point condition
\eqref{DEQ}, $\mathbf{h}^{\star}$ should satisfy 
\begin{align}
\Vert\mathbf{h}_{\mathrm{G}}^{\star}\Vert_{2}= & \Vert R_{\mathbf{\Theta}}\left(\eta\mathbf{u}+(\mathbf{I}_{2N}-\eta\mathbf{A}^{T}\mathbf{C}^{-1}\mathbf{A})\mathbf{h}^{\star}\right)-R_{\mathbf{\Theta}}(\bm{0})\Vert_{2}\nonumber \\
\le & L\Vert\eta\mathbf{u}+(\mathbf{I}_{2N}-\eta\mathbf{A}^{T}\mathbf{\mathbf{C}^{-1}A})\mathbf{h}^{\star}\Vert_{2}\quad\left(\text{recall \eqref{contraction}}\right)\nonumber \\
= & L\Vert\eta\mathbf{A}^{T}\mathbf{\mathbf{C}^{-1}A}\left(\mathbf{Ph}-\mathbf{P}\mathbf{h}_{\mathrm{G}}^{\star}+\mathbf{A}^{T}\mathbf{n}\right)+\mathbf{h}_{\mathrm{G}}^{\star}\Vert_{2}\nonumber \\
\le & L\Vert\mathbf{h}_{\mathrm{G}}^{\star}\Vert_{2}+\eta L\cdot\Vert\mathbf{\mathbf{C}}^{-1}\Vert_{2}\cdot\left(\Vert\mathbf{P}\mathbf{h}_{\mathrm{G}}^{\star}-\mathbf{Ph}\Vert_{2}+\Vert\mathbf{n}\Vert_{2}\right).
\end{align}
Let $\gamma=\eta L/(1-L)$, and recall Assumptions \ref{good_pmse}
and \ref{psnr bound}, then 
\begin{align}
\Vert\mathbf{h}_{\mathrm{G}}^{\star}\Vert_{2}\le & \gamma\cdot\Vert\mathbf{\mathbf{C}}^{-1}\Vert_{2}\cdot\left(\Vert\mathbf{P}\mathbf{h}_{\mathrm{G}}^{\star}-\mathbf{Ph}\Vert_{2}+\Vert\mathbf{n}\Vert_{2}\right)\nonumber \\
\le & \gamma\cdot\Vert\mathbf{\mathbf{C}}^{-1}\Vert_{2}\cdot\left(\beta+\xi\right)^{2}\cdot\Vert\mathbf{Ph}\Vert_{2}.
\end{align}
It follows that 
\begin{equation}
\sqrt{1-\frac{\Vert\mathbf{P}\mathbf{h}_{\mathrm{G}}^{\star}\Vert_{2}^{2}}{\Vert\mathbf{h}_{\mathrm{G}}^{\star}\Vert_{2}^{2}}}\le\sqrt{1-\frac{\left(1-\beta\right)^{2}}{\gamma^{2}\Vert\mathbf{\mathbf{C}}^{-1}\Vert_{2}^{2}\left(\beta+\xi\right)^{2}}}.\label{unhandled_2}
\end{equation}
For $\frac{\Vert\mathbf{P}\mathbf{h}^{\star}\Vert_{1}}{\Vert\mathbf{P}\mathbf{h}^{\star}\Vert_{2}}$,
we recall Assumption \ref{good_pmse} again and obtain that
\begin{align}
\frac{\Vert\mathbf{P}\mathbf{h}_{\mathrm{G}}^{\star}\Vert_{1}}{\Vert\mathbf{P}\mathbf{h}_{\mathrm{G}}^{\star}\Vert_{2}}\le & \frac{\Vert\mathbf{Ph}+\mathbf{P}\mathbf{h}_{\mathrm{G}}^{\star}-\mathbf{Ph}\Vert_{1}}{\Vert\mathbf{P}\mathbf{h}_{\mathrm{G}}^{\star}\Vert_{2}}\nonumber \\
\le & \frac{1}{1-\beta}\left(\frac{\Vert\mathbf{Ph}\Vert_{1}}{\Vert\mathbf{Ph}\Vert_{2}}+\sqrt{2N}\frac{\Vert\mathbf{P}\mathbf{h}_{\mathrm{G}}^{\star}-\mathbf{Ph}\Vert_{2}}{\Vert\mathbf{Ph}\Vert_{2}}\right)\nonumber \\
\le & \frac{1}{1-\beta}\left(\frac{\Vert\mathbf{Ph}\Vert_{1}}{\Vert\mathbf{Ph}\Vert_{2}}+\beta\sqrt{2N}\right).\label{unhandled_3}
\end{align}
We shorten the notation $\sqrt{\frac{N}{M}}\mathbf{A}$ to $\tilde{\mathbf{A}}$,
which leads to 
\begin{equation}
\frac{\Vert\mathbf{Ph}\Vert_{1}}{\Vert\mathbf{Ph}\Vert_{2}}=\frac{\frac{N}{M}\Vert\mathbf{A}^{T}\mathbf{A}\mathbf{h}\Vert_{1}}{\frac{N}{M}\Vert\mathbf{A}^{T}\mathbf{A}\mathbf{h}\Vert_{2}}=\frac{\Vert\tilde{\mathbf{A}}^{T}\tilde{\mathbf{A}}\mathbf{h}\Vert_{1}}{\Vert\tilde{\mathbf{A}}^{T}\tilde{\mathbf{A}}\mathbf{h}\Vert_{2}}.\label{unhandled}
\end{equation}
Define $S=\mathrm{supp}(\mathbf{h})$ and $\vert S\vert\le2k$. Define
the submatrix $\tilde{\mathbf{A}}_{S}\in\mathbb{R}^{2M\times2N}$
as 
\begin{equation}
\tilde{\mathbf{A}}_{S}(i)=\begin{cases}
\tilde{\mathbf{A}}(i), & {\text{if}}\ i\in S\\
\bm{0}, & {\text{if}}\ i\notin S
\end{cases}\label{submatrix}
\end{equation}
where $\tilde{\mathbf{A}}(i)$ denotes the $i$-th column vector of
$\tilde{\mathbf{A}}$, and $\tilde{\mathbf{A}}_{S^{c}}=\tilde{\mathbf{A}}-\tilde{\mathbf{A}}_{S}$.
Decomposing the numerator of \eqref{unhandled}, we have 
\begin{align}
\Vert\tilde{\mathbf{A}}^{T}\tilde{\mathbf{A}}\mathbf{h}\Vert_{1}= & \Big\Vert\left(\tilde{\mathbf{A}}_{S}+\tilde{\mathbf{A}}_{S^{c}}\right)^{T}\tilde{\mathbf{A}}\mathbf{h}\Big\Vert_{1}\nonumber \\
\le & \Vert\mathbf{h}\Vert_{1}+\Big\Vert\left(\tilde{\mathbf{A}}_{S}+\tilde{\mathbf{A}}_{S^{c}}\right)^{T}\tilde{\mathbf{A}}\mathbf{h}-\mathbf{h}\Big\Vert_{1}\nonumber \\
\le & \Vert\mathbf{h}\Vert_{1}+\Vert\tilde{\mathbf{A}}_{S}^{T}\tilde{\mathbf{A}}\mathbf{h}-\mathbf{h}\Vert_{1}+\Vert\tilde{\mathbf{A}}_{S^{c}}^{T}\tilde{\mathbf{A}}_{S}\mathbf{h}\Vert_{1}.\label{decomposed}
\end{align}
The second term in \eqref{decomposed} can be bounded as
\begin{align}
\Vert\tilde{\mathbf{A}}_{S}^{T}\tilde{\mathbf{A}}\mathbf{h}-\mathbf{h}\Vert_{1}= & \Vert\tilde{\mathbf{A}}_{S}^{T}\tilde{\mathbf{A}}_{S}-\mathbf{I}_{S}\Vert_{1}\Vert\mathbf{h}\Vert_{1}\nonumber \\
\le & \sqrt{2k}\Vert\tilde{\mathbf{A}}_{S}^{T}\tilde{\mathbf{A}}_{S}-\mathbf{I}_{S}\Vert_{2}\Vert\mathbf{h}\Vert_{1}\nonumber \\
\le & \sqrt{2k}\cdot\delta_{2k}\Vert\mathbf{h}\Vert_{1},
\end{align}
where $\mathbf{I}_{S}$ denotes the submatrix of $\mathbf{I}_{2N}$
by the definition \eqref{submatrix}. According to Proposition 6.2
in \cite{Foucart_2013}, for $\forall i\neq j$, 
\begin{equation}
\vert\tilde{\mathbf{A}}(j)^{T}\tilde{\mathbf{A}}(i)\vert\le\delta_{2},
\end{equation}
where $\delta_{2}$ is 2nd-order RIP constant of $\tilde{\mathbf{A}}$.
Then the last term in \eqref{decomposed} can be bounded as
\begin{align}
\Vert\tilde{\mathbf{A}}_{S^{c}}^{T}\tilde{\mathbf{A}}_{S}\mathbf{h}\Vert_{1}\le & \sum_{j\notin S}\left|\tilde{\mathbf{A}}(j)^{T}\left[\sum_{i\in S}\tilde{\mathbf{A}}(i)\mathbf{h}(i)\right]\right|\nonumber \\
\le & (2N-2k)\cdot2k\delta_{2}\cdot\Vert\mathbf{h}\Vert_{1}.\label{eq: bound_numerator}
\end{align}
Then \eqref{decomposed} can be further bounded as 
\begin{equation}
\Vert\tilde{\mathbf{A}}^{T}\tilde{\mathbf{A}}\mathbf{h}\Vert_{1}\le\left(1+\left[\sqrt{2k}\cdot\delta_{2k}+(2N-2k)\cdot2k\delta_{2}\right]\right)\Vert\mathbf{h}\Vert_{1}.
\end{equation}
For the denominator of \eqref{unhandled}, the following lower bound
can be obtained: 
\begin{align}
\Vert\tilde{\mathbf{A}}^{T}\tilde{\mathbf{A}}\mathbf{h}\Vert_{2}= & \left\Vert \left(\tilde{\mathbf{A}}_{S}+\tilde{\mathbf{A}}_{S^{c}}\right)^{T}\tilde{\mathbf{A}}_{S}\mathbf{h}\right\Vert _{2}\nonumber \\
\ge & \left\Vert \tilde{\mathbf{A}}_{S}^{T}\tilde{\mathbf{A}}_{S}\mathbf{h}\right\Vert _{2}\ge(1-\delta_{2k})\Vert\mathbf{h}\Vert_{2}.\label{eq: bound_denominator}
\end{align}
Combining \eqref{eq: bound_numerator} and \eqref{eq: bound_denominator},
\eqref{unhandled} can be bounded as 
\begin{equation}
\frac{\Vert\mathbf{Ph}\Vert_{1}}{\Vert\mathbf{Ph}\Vert_{2}}\le\frac{1+\left[\sqrt{2k}\cdot\delta_{2k}+(2N-2k)\cdot2k\delta_{2}\right]}{1-\delta_{2k}}\frac{\Vert\mathbf{h}\Vert_{1}}{\Vert\mathbf{h}\Vert_{2}}.\label{unhandled_4}
\end{equation}
Combining \eqref{unhandled_1}, \eqref{unhandled_2}, \eqref{unhandled_3}
and \eqref{unhandled_4}, we obtain
\begin{align}
\ell_{1/2}(\mathbf{h}_{\mathrm{G}}^{\star})\le & \frac{1+\left[\sqrt{2k}\cdot\delta_{2k}+(2N-2k)\cdot2k\delta_{2}\right]}{(1-\beta)(1-\delta_{2k})}\cdot\sqrt{2k}\nonumber \\
 & +\sqrt{2N}\left(\frac{\beta}{1-\beta}+\sqrt{1-\frac{\left(1-\beta\right)^{2}}{\gamma^{2}\Vert\mathbf{\mathbf{C}}^{-1}\Vert_{2}^{2}\left(\beta+\xi\right)^{2}}}\right).
\end{align}
For $\mathbf{h}_{\mathrm{ora}}^{\star}$, recalling Assumption \ref{good_pmse},
we have
\begin{align}
\Vert\mathbf{h}_{\mathrm{ora}}^{\star}\Vert_{2}= & \Vert\mathbf{h}+\mathbf{h}_{\mathrm{ora}}^{\star}-\mathbf{h}\Vert_{2}\nonumber \\
\ge & \big\vert\Vert\mathbf{h}\Vert_{2}-\Vert\mathbf{h}_{\mathrm{ora}}^{\star}-\mathbf{h}\Vert_{2}\big\vert\ge(1-\omega)\Vert\mathbf{h}\Vert_{2}.
\end{align}
Define $\mathcal{I}=\mathrm{\mathrm{supp}(\mathbf{h}_{\mathrm{ora}}^{\star})\cap\mathrm{supp}(\mathbf{h})}$,
and let $\mathbf{h}_{\mathrm{ora},\mathcal{I}}^{\star}$ be the vector
that keeps the values of $\mathbf{h}_{\mathrm{ora}}^{\star}$ on $\mathcal{I}$
and sets the remaining entries to zero. Define $\mathbf{h}_{\mathrm{ora},\mathcal{I}^{c}}^{\star}=\mathbf{h}_{\mathrm{ora}}^{\star}-\mathbf{h}_{\mathrm{ora},\mathcal{I}}^{\star}$.
Recalling Assumption \ref{good_pmse} again and using the fact that
$\Vert\mathbf{h}_{\mathrm{ora},\mathcal{I}^{c}}^{\star}\Vert_{1}/\sqrt{2N-2k}\le\Vert\mathbf{h}_{\mathrm{ora},\mathcal{I}^{c}}^{\star}\Vert_{2}\le\Vert\mathbf{h}_{\mathrm{ora}}^{\star}-\mathbf{h}\Vert_{2}\le\omega\Vert\mathbf{h}\Vert_{2}$,
we obtain the following upper bound for $\ell_{1/2}(\mathbf{h}_{\mathrm{ora}}^{\star})$
:
\begin{equation}
\begin{array}{rl}
\ell_{1/2}(\mathbf{h}_{\mathrm{ora}}^{\star}) & ={\displaystyle \frac{\Vert\mathbf{h}_{\mathrm{ora}}^{\star}\Vert_{1}}{\Vert\mathbf{h}_{\mathrm{ora}}^{\star}\Vert_{2}}}\\
 & \le{\displaystyle \frac{1}{1-\omega}\cdot\left(\frac{\Vert\mathbf{h}_{\mathrm{ora},\mathcal{I}}^{\star}\Vert_{1}}{\Vert\mathbf{h}\Vert_{2}}+{\displaystyle \frac{\Vert\mathbf{h}_{\mathrm{ora},\mathcal{I}^{c}}^{\star}\Vert_{1}}{\Vert\mathbf{h}\Vert_{2}}}\right)}\\
 & \le{\displaystyle \frac{1}{1-\omega}\cdot\left(\sqrt{2k}+\sqrt{2N-2k}\cdot\omega\right)}.
\end{array}
\end{equation}

\section{Proof of Theorem \ref{MSE bound of h_G and h_ora}}

\label{proof of MSE bound of h_G and h_ora}

For brevity, we denote $\tilde{\mathbf{A}}=\sqrt{\frac{N}{M}}\mathbf{A}$.
We only show the proof of \eqref{MSE_le_GSURE}, since \eqref{MSE_ge_GSURE}
follows a similar process. Notice that $\mathrm{supp}(\mathbf{h})\subseteq\mathrm{supp}(\mathbf{h}_{T}^{\star})$,
which means $\mathrm{supp}(\mathbf{h})\cap\mathrm{supp}(\mathbf{h}_{T^{c}}^{\star})=\varnothing$,
then 
\begin{align}
\Vert\mathbf{h}^{\star}-\mathbf{h}\Vert_{2}^{2}= & \Vert\mathbf{h}_{T}^{\star}-\mathbf{h}\Vert_{2}^{2}+\Vert\mathbf{h}_{T^{c}}^{\star}\Vert_{2}^{2},\label{h_ht}\\
\Big\Vert\mathbf{A}\left(\mathbf{h}^{\star}-\mathbf{h}\right)\Big\Vert_{2}^{2}= & \Vert\mathbf{A}(\mathbf{h}_{T}^{\star}-\mathbf{h})\Vert_{2}^{2}+\Vert\mathbf{A}\mathbf{h}_{T^{c}}^{\star}\Vert_{2}^{2}\nonumber \\
 & +2\langle\mathbf{A}(\mathbf{h}_{T}^{\star}-\mathbf{h}),\mathbf{A}\mathbf{h}_{T^{c}}^{\star}\rangle.\label{Ah_Ahs}
\end{align}
Recalling the RIP property for a partially orthogonal matrix $\mathbf{A}\in\mathbb{R}^{2M\times2N}$,
and supposing that $\Vert\mathbf{A}\Vert_{\infty}\le\zeta/\sqrt{2N}$,
then according to \cite{haviv2017restricted}, if 
\begin{equation}
2M\ge\frac{C\zeta^{2}}{\delta^{2}}\cdot T\cdot\log^{2}\left(\frac{T}{\delta}\right)\cdot\log^{2}\left(\frac{1}{\delta}\right)\cdot\log\left(2N\right),
\end{equation}
then, with high probability, matrix $\tilde{\mathbf{A}}$ satisfies
the RIP of order $T$ with $\delta_{T}\le\delta$, i.e.,
\begin{equation}
\left(1-\delta_{T}\right)\Vert\mathbf{h}_{T}^{\star}-\mathbf{h}\Vert_{2}^{2}\le\left\Vert \tilde{\mathbf{A}}\left(\mathbf{h}_{T}^{\star}-\mathbf{h}\right)\right\Vert _{2}^{2}\le(1+\delta_{T})\Vert\mathbf{h}_{T}^{\star}-\mathbf{h}\Vert_{2}^{2}.\label{MSE_PMSE}
\end{equation}
 Combining \eqref{h_ht}-\eqref{Ah_Ahs} with the left-hand side of
\eqref{MSE_PMSE}, 
\begin{align}
\left(1-\delta_{T}\right)\Vert\mathbf{h}^{\star}-\mathbf{h}\Vert_{2}^{2}\le & \left\Vert \tilde{\mathbf{A}}\left(\mathbf{h}^{\star}-\mathbf{h}\right)\right\Vert _{2}^{2}+\left(1-\delta_{T}\right)\Vert\mathbf{h}_{T^{c}}^{\star}\Vert_{2}^{2}\nonumber \\
 & -\Vert\tilde{\mathbf{A}}\mathbf{h}_{T^{c}}^{\star}\Vert_{2}^{2}-2\langle\tilde{\mathbf{A}}(\mathbf{h}_{T}^{\star}-\mathbf{h}),\tilde{\mathbf{A}}\mathbf{h}_{T^{c}}^{\star}\rangle.\label{combined}
\end{align}
We decompose $\mathbf{h}_{T^{c}}^{\star}=\mathbf{h}_{T_{1}^{c}}^{\star}+\mathbf{h}_{T_{2}^{c}}^{\star}+\dots$,
where $T_{i}^{c}$ is the index set corresponding to the $i$-th $T$
largest entries of $\mathbf{h}_{T^{c}}^{\star}$. Using the fact (\cite{wright2022high},
Lemma 3.16) that 
\begin{equation}
\left|\langle\tilde{\mathbf{A}}\mathbf{h}_{T_{i}^{c}}^{\star},\tilde{\mathbf{A}}\mathbf{h}_{T_{j}^{c}}^{\star}\rangle\right|\le\delta_{2T}\cdot\Vert\mathbf{h}_{T_{i}^{c}}^{\star}\Vert_{2}\Vert\mathbf{h}_{T_{j}^{c}}^{\star}\Vert_{2},\label{fact1}
\end{equation}
the last term in \eqref{combined} can be bounded through 
\begin{align}
\left|\langle\tilde{\mathbf{A}}(\mathbf{h}_{T}^{\star}-\mathbf{h}),\tilde{\mathbf{A}}\mathbf{h}_{T^{c}}^{\star}\rangle\right|\le & \sum_{j\ge1}\left|\langle\tilde{\mathbf{A}}(\mathbf{h}_{T}^{\star}-\mathbf{h}),\tilde{\mathbf{A}}\mathbf{h}_{T_{j}^{c}}^{\star}\rangle\right|\nonumber \\
\le & \delta_{2T}\Vert\mathbf{h}_{T}^{\star}-\mathbf{h}\Vert_{2}\left(\sum_{j\ge1}\Vert\mathbf{h}_{T_{j}^{c}}^{\star}\Vert_{2}\right)\nonumber \\
\le & \frac{1}{2}\delta_{2T}\sqrt{\frac{2N}{T}}\Vert\mathbf{h}^{\star}-\mathbf{h}\Vert_{2}^{2}.
\end{align}
The third term in \eqref{combined} can be bounded through RIP and
\eqref{fact1}, 
\begin{align}
\Vert\tilde{\mathbf{A}}\mathbf{h}_{T^{c}}^{\star}\Vert_{2}^{2}= & \sum_{j\ge1}\left\Vert \tilde{\mathbf{A}}\mathbf{h}_{T_{j}^{c}}^{\star}\right\Vert _{2}^{2}+\sum_{i\neq j}\left|\langle\tilde{\mathbf{A}}\mathbf{h}_{T_{i}^{c}}^{\star},\tilde{\mathbf{A}}\mathbf{h}_{T_{j}^{c}}^{\star}\rangle\right|\nonumber \\
\ge & (1-\delta_{T})\cdot\sum_{j\ge1}\Vert\mathbf{h}_{T_{j}^{c}}^{\star}\Vert_{2}^{2}-\delta_{2T}\cdot\sum_{i\neq j}\Vert\mathbf{h}_{T_{i}^{c}}^{\star}\Vert_{2}\Vert\mathbf{h}_{T_{j}^{c}}^{\star}\Vert_{2}\nonumber \\
\ge & (1-\delta_{T})\Vert\mathbf{h}_{T^{c}}^{\star}\Vert_{2}^{2}-\delta_{2T}\frac{2N-T}{T}\Vert\mathbf{h}_{T^{c}}^{\star}\Vert_{2}^{2}.
\end{align}
Substituting into \eqref{combined}, we obtain
\begin{alignat}{1}
\left(1-\delta_{T}\right)\Vert\mathbf{h}^{\star}-\mathbf{h}\Vert_{2}^{2}\le & \Big\Vert\tilde{\mathbf{A}}\left(\mathbf{h}^{\star}-\mathbf{h}\right)\Big\Vert_{2}^{2}+\delta_{2T}\frac{2N-T}{T}\Vert\mathbf{h}_{T^{c}}^{\star}\Vert_{2}^{2}\nonumber \\
 & +\delta_{2T}\cdot\sqrt{\frac{2N}{T}}\Vert\mathbf{h}^{\star}-\mathbf{h}\Vert_{2}^{2}.
\end{alignat}
If there exists a constant $C$ such that
\begin{equation}
M\ge\frac{C\zeta^{2}}{\delta^{2}}\cdot T\cdot\log^{2}\left(\frac{2T}{\delta}\right)\cdot\log\left(\frac{N}{T}\right),
\end{equation}
then the $2T$-order RIP constant of the matrix $\tilde{\mathbf{A}}$
satisfies $\delta_{2T}\le\delta$ with high probability. By choosing
a suitable $\delta$ such that $\epsilon=\delta+\sqrt{\frac{2N}{T}}\cdot\delta\ll1$
holds with high probability, we obtain that, with high probability,
\begin{align}
\left(1-\epsilon\right)\Vert\mathbf{h}^{\star}-\mathbf{h}\Vert_{2}^{2}\le & \frac{N}{M}\Big\Vert\mathbf{A}\left(\mathbf{h}^{\star}-\mathbf{h}\right)\Big\Vert_{2}^{2}+\delta\cdot\frac{2N-T}{T}\Vert\mathbf{h}_{T^{c}}^{\star}\Vert_{2}^{2}\nonumber \\
= & \frac{N}{M}\Big\Vert\mathbf{P}\left(\mathbf{h}^{\star}-\mathbf{h}\right)\Big\Vert_{2}^{2}+\delta\cdot\frac{2N-T}{T}\Vert\mathbf{h}_{T^{c}}^{\star}\Vert_{2}^{2}.
\end{align}
Taking the mathematical expectation, and recalling \eqref{PMSE},
the following holds:
\begin{align}
\left(1-\epsilon\right)\mathrm{MSE}_{\mathrm{G}}\le & \frac{N}{M}\mathrm{PMSE}(\mathbf{h}_{\mathrm{G}}^{\star})+\delta\cdot\frac{2N-T}{T}\mathbb{E}\Vert\mathbf{h}_{\mathrm{G},T^{c}}^{\star}\Vert_{2}^{2}\nonumber \\
 & \frac{N}{M}\mathrm{GSURE}(\mathbf{h}_{\mathrm{G}}^{\star})+\delta\cdot\frac{2N-T}{T}\mathbb{E}\Vert\mathbf{h}_{\mathrm{G},T^{c}}^{\star}\Vert_{2}^{2}.
\end{align}
Similarly, combining \eqref{h_ht}-\eqref{Ah_Ahs} with the right-hand
side of \eqref{MSE_PMSE} and substituting $\mathbf{h}_{\mathrm{ora}}^{\star}$,
we can prove \eqref{MSE_ge_GSURE}.

\section{Proof of Theorem \ref{oracle inequality}}

\label{proof of oracle inequality} We give a lemma about DEQ in advance,
which is very useful in the subsequent Theorems.

\begin{lemma}\label{Lip_bound} For the DEQ defined in \eqref{DEQ},
if $f_{\mathbf{\Theta}}(\cdot;\mathbf{u}):\mathbb{R}^{2N}\to\mathbb{R}^{2N}$
is an $L$-Lipschitz contraction with $L\in(0,1)$ and Assumption
\ref{zero_output} holds, then 
\begin{equation}
\Vert\mathbf{h}_{\mathbf{\Theta}}^{\star}(\mathbf{u})\Vert_{2}\le\frac{\eta L}{1-L}\Vert\mathbf{u}\Vert_{2}=\gamma\cdot\Vert\mathbf{u}\Vert_{2}.
\end{equation}

\label{proof_Lip_bound}

\end{lemma}
\begin{IEEEproof}
Let $\mathbf{r}^{(k)}=\eta\mathbf{u}+(\mathbf{I}-\eta\mathbf{A}^{T}\mathbf{C}^{-1}\mathbf{A})\mathbf{h}_{\mathbf{\Theta}}^{(k-1)}(\mathbf{u})$.
For any $\mathbf{u},\mathbf{u}_{1}\in\mathbb{R}^{2N}$, at the $k$-th
iteration, it holds that
\begin{align}
 & \Vert\mathbf{h}_{\mathbf{\Theta}}^{(k)}(\mathbf{u}_{2})-\mathbf{h}_{\mathbf{\Theta}}^{(k)}(\mathbf{u}_{1})\Vert_{2}\nonumber \\
\le & \Big\Vert R_{\mathbf{\Theta}}\left(\mathbf{r}_{2}^{(k)}\right)-R_{\mathbf{\Theta}}\left(\mathbf{r}_{1}^{(k)}\right)\Big\Vert_{2}\nonumber \\
\le & L_{2}\cdot\big\Vert\mathbf{r}_{2}^{(k)}-\mathbf{r}_{1}^{(k)}\big\Vert_{2}\nonumber \\
\le & \eta L\cdot\Vert\mathbf{u}_{2}-\mathbf{u}_{1}\Vert_{2}+L\cdot\Vert\mathbf{h}_{\mathbf{\Theta}}^{(k-1)}(\mathbf{u}_{2})-\mathbf{h}_{\mathbf{\Theta}}^{(k-1)}(\mathbf{u}_{1})\Vert_{2}\nonumber \\
\le & \eta L\cdot\Vert\mathbf{u}_{2}-\mathbf{u}_{1}\Vert_{2}\nonumber \\
 & +L\Big(L\Vert\mathbf{h}_{\mathbf{\Theta}}^{(k-2)}(\mathbf{u}_{2})-\mathbf{h}_{\mathbf{\Theta}}^{(k-2)}(\mathbf{u}_{1})\Vert_{2}+\eta L\Vert\mathbf{u}_{2}-\mathbf{u}_{1}\Vert_{2}\Big)\nonumber \\
\le & \dots\nonumber \\
\le & L^{k}\cdot\Vert\mathbf{h}_{\mathbf{\Theta}}^{(0)}(\mathbf{u}_{2})-\mathbf{h}_{\mathbf{\Theta}}^{(0)}(\mathbf{u}_{1})\Vert_{2}+\eta L\Vert\mathbf{u}_{2}-\mathbf{u}_{1}\Vert_{2}\cdot\sum_{i=0}^{k-1}L^{i}.\label{u2_u1}
\end{align}
The first term in \eqref{u2_u1} converges to $0$ as $k\to\infty$.
Then we have
\begin{equation}
\Vert\mathbf{h}_{\mathbf{\Theta}}^{(k)}(\mathbf{u}_{2})-\mathbf{h}_{\mathbf{\Theta}}^{(k)}(\mathbf{u}_{1})\Vert_{2}\le\frac{\eta L\left(1-L^{k}\right)}{1-L}\cdot\Vert\mathbf{u}_{2}-\mathbf{u}_{1}\Vert_{2}.
\end{equation}
Then 
\begin{align}
\Vert\mathbf{h}_{\mathbf{\Theta}}^{\star}(\mathbf{u}_{2})-\mathbf{h}_{\mathbf{\Theta}}^{\star}(\mathbf{u}_{1})\Vert_{2}= & \lim_{k\to\infty}\Vert\mathbf{h}_{\mathbf{\Theta}}^{(k)}(\mathbf{u}_{2})-\mathbf{h}_{\mathbf{\Theta}}^{(k)}(\mathbf{u}_{1})\Vert_{2}\nonumber \\
\le & \gamma\cdot\Vert\mathbf{u}_{2}-\mathbf{u}_{1}\Vert_{2}.
\end{align}
Let $\mathbf{u}_{2}=\mathbf{u}$ and $\mathbf{u}_{1}=\bm{0}$. Using
Assumption \ref{zero_output}, we have
\begin{equation}
\Vert\mathbf{h}_{\mathbf{\Theta}}^{\star}(\mathbf{u})\Vert_{2}\le\gamma\cdot\Vert\mathbf{u}\Vert_{2}.
\end{equation}
\end{IEEEproof}
Now we provide the proof of Theorem \ref{oracle inequality}.

According to Theorem \ref{MSE bound of h_G and h_ora}, for $\mathbf{h}_{\mathrm{G}}^{\star}$
and $\mathbf{h}_{\mathrm{ora}}^{\star}$, we can gain, respectively,
\begin{equation}
\mathrm{MSE}_{\mathrm{G}}\le\frac{N}{M}\frac{\mathrm{GSURE}(\mathbf{h}_{\mathrm{G}}^{\star})}{1-\epsilon}+\frac{\delta}{1-\epsilon}\frac{2N-T}{T}\mathbb{E}\Vert\mathbf{h}_{\mathrm{G},\mathcal{T}^{c}}^{\star}\Vert_{2}^{2},\label{MSE_le_GSURE_Theta_G}
\end{equation}
\begin{equation}
\mathrm{MSE}_{\mathrm{ora}}\ge\frac{N}{M}\frac{\mathrm{GSURE}(\mathbf{h}_{\mathrm{ora}}^{\star})}{1+\epsilon}-\frac{\delta}{1+\epsilon}\frac{2N-T}{T}\mathbb{E}\Vert\mathbf{h}_{\mathrm{ora},\mathcal{R}^{c}}^{\star}\Vert_{2}^{2}.\label{MSE_ge_GSURE_Theta_MSE}
\end{equation}
First, since $s\le T$, it follows that $\Vert\mathbf{h}_{\mathrm{G},\mathcal{T}^{c}}^{\star}\Vert_{2}^{2}\le\Vert\mathbf{h}_{\mathrm{G},s^{c}}^{\star}\Vert_{2}^{2}$
and $\Vert\mathbf{h}_{\mathrm{ora},\mathcal{R}^{c}}^{\star}\Vert_{2}^{2}\le\Vert\mathbf{h}_{\mathrm{ora},s^{c}}^{\star}\Vert_{2}^{2}$.
Second, according to Assumption \ref{psnr bound}, $\Vert\mathbf{n}\Vert_{2}/\Vert\mathbf{Ah}\Vert_{2}\le\xi$
holds with high probability. Finally, according to Lemma \ref{Lip_bound},
we have $\|\mathbf{u}\|_{2}^{2}=\|\mathbf{Ph}+\mathbf{A}^{T}\mathbf{n}\|_{2}^{2}\le2\left(\|\mathbf{Ah}\|_{2}^{2}+\|\mathbf{n}\|_{2}^{2}\right)\le2(1+\xi^{2})\Vert\mathbf{Ah}\Vert_{2}^{2}$.
Therefore, the following inequality holds with high probability:
\begin{align}
\mathbb{E}\Vert\mathbf{h}_{\mathrm{G},\mathcal{T}^{c}}^{\star}\Vert_{2}^{2}= & \mathbb{E}\left[\frac{\Vert\mathbf{h}_{\mathrm{G},\mathcal{T}^{c}}^{\star}\Vert_{2}^{2}}{\Vert\mathbf{h}_{\mathrm{G}}^{\star}\Vert_{2}^{2}}\cdot\Vert\mathbf{h}_{G}^{\star}\Vert_{2}^{2}\right]\nonumber \\
\le & \gamma^{2}\mathbb{E}\left[\frac{\Vert\mathbf{h}_{\mathrm{G},s^{c}}^{\star}\Vert_{2}^{2}}{\Vert\mathbf{h}_{\mathrm{G}}^{\star}\Vert_{2}^{2}}\cdot\Vert\mathbf{u}\Vert_{2}^{2}\right]\nonumber \\
\le & 2(1+\xi^{2})\gamma^{2}\mathbb{E}\left[\frac{\Vert\mathbf{h}_{\mathrm{G},s^{c}}^{\star}\Vert_{2}^{2}}{\Vert\mathbf{h}_{\mathrm{G}}^{\star}\Vert_{2}^{2}}\cdot\Vert\mathbf{Ah}\Vert_{2}^{2}\right].
\end{align}
Recalling Theorem $\ref{SGF Upper Bound}$, $\ell_{1/2}(\mathbf{h}_{\mathrm{G}}^{\star})$
decreases monotonically as $\Vert\mathbf{Ah}\Vert_{2}^{2}$ increases.
Additionally, condition \eqref{T_le_N} implies that $T<N$. Hence,
when $\ell_{1/2}(\mathbf{h}_{\mathrm{G}}^{\star})\in(0,\sqrt{T}]$,
$\frac{\Vert\mathbf{h}_{\mathrm{G},s^{c}}^{\star}\Vert_{1}}{\Vert\mathbf{h}^{\star}\Vert_{2}}$
increases monotonically as $\ell_{1/2}(\mathbf{h}_{\mathrm{G}}^{\star})$
increases. This indicates a negative correlation between the two,
which can be further analyzed by covariance analysis:
\begin{align}
 & \mathrm{cov}\left(\frac{\Vert\mathbf{h}_{\mathrm{G},s^{c}}^{\star}\Vert_{2}^{2}}{\Vert\mathbf{h}_{\mathrm{G}}^{\star}\Vert_{2}^{2}},\Vert\mathbf{Ah}\Vert_{2}^{2}\right)\nonumber \\
= & \mathbb{E}\left[\frac{\Vert\mathbf{h}_{\mathrm{G},s^{c}}^{\star}\Vert_{2}^{2}}{\Vert\mathbf{h}_{\mathrm{G}}^{\star}\Vert_{2}^{2}}\cdot\Vert\mathbf{Ah}\Vert_{2}^{2}\right]-\mathbb{E}\left[\frac{\Vert\mathbf{h}_{\mathrm{G},s^{c}}^{\star}\Vert_{2}^{2}}{\Vert\mathbf{h}^{\star}\Vert_{2}^{2}}\right]\cdot\mathbb{E}\Vert\mathbf{\mathbf{Ah}}\Vert_{2}^{2}\nonumber \\
\le & 0.
\end{align}
\begin{equation}
\Rightarrow\mathbb{E}\left[\frac{\Vert\mathbf{h}_{\mathrm{G},s^{c}}^{\star}\Vert_{2}^{2}}{\Vert\mathbf{h}_{\mathrm{G}}^{\star}\Vert_{2}^{2}}\cdot\Vert\mathbf{Ph}\Vert_{2}^{2}\right]\le\mathbb{E}\left[\frac{\Vert\mathbf{h}_{\mathrm{G},s^{c}}^{\star}\Vert_{2}^{2}}{\Vert\mathbf{h}_{\mathrm{G}}^{\star}\Vert_{2}^{2}}\right]\cdot\mathbb{E}\Vert\mathbf{Ah}\Vert_{2}^{2}.\label{eq: split_E}
\end{equation}
Note that \eqref{eq: split_E} also holds for $\mathbf{h}_{\mathrm{ora}}^{\star}$.
Recalling Theorem \ref{SGF Upper Bound}-\ref{bounded l_1/2} and
noting that SGF is a concave function, we have 
\begin{equation}
\mathbb{E}\left[\frac{\Vert\mathbf{h}_{\mathrm{G},s^{c}}^{\star}\Vert_{2}^{2}}{\Vert\mathbf{h}_{\mathrm{G}}^{\star}\Vert_{2}^{2}}\right]\le g(s_{\mathrm{G}}),\quad\mathbb{E}\left[\frac{\Vert\mathbf{h}_{\mathrm{ora},s^{c}}^{\star}\Vert_{2}^{2}}{\Vert\mathbf{h}_{\mathrm{ora}}^{\star}\Vert_{2}^{2}}\right]\le g(s_{\mathrm{ora}}),
\end{equation}
here $s_{\mathrm{G}}$ and $s_{\mathrm{ora}}$ are defined in \eqref{s_G}
and \eqref{s_ora}, respectively. If we let
\begin{equation}
s=\max\left\{ \left\lceil s_{\mathrm{G}}\right\rceil ,\left\lceil s_{\mathrm{ora}}\right\rceil \right\} ,
\end{equation}
then it follows that 
\begin{equation}
\max\left\{ \mathbb{E}\left[\frac{\Vert\mathbf{h}_{\mathrm{G},s^{c}}^{\star}\Vert_{2}^{2}}{\Vert\mathbf{h}_{\mathrm{G}}^{\star}\Vert_{2}^{2}}\right],\mathbb{E}\left[\frac{\Vert\mathbf{h}_{\mathrm{ora},s^{c}}^{\star}\Vert_{2}^{2}}{\Vert\mathbf{h}_{\mathrm{ora}}^{\star}\Vert_{2}^{2}}\right]\right\} \leq g(s).
\end{equation}
It consequently follows that
\begin{align}
\mathbb{E}\Vert\mathbf{h}_{\mathrm{G},\mathcal{T}^{c}}^{\star}\Vert_{2}^{2}\le & 2(1+\xi^{2})\gamma^{2}\mathbb{E}\left[\frac{\Vert\mathbf{h}_{\mathrm{G},s^{c}}^{\star}\Vert_{2}^{2}}{\Vert\mathbf{h}_{\mathrm{G}}^{\star}\Vert_{2}^{2}}\right]\cdot\mathbb{E}\Vert\mathbf{Ah}\Vert_{2}^{2}\nonumber \\
\le & 2(1+\xi^{2})\gamma^{2}\cdot g(s)\cdot\mathbb{E}\Vert\mathbf{Ah}\Vert_{2}^{2}.
\end{align}
Similarly, the following upper bound of the residual term of the $T$-largest
sparse approximation of the DEQ oracle estimate also holds with high
probability:
\begin{equation}
\mathbb{E}\Vert\mathbf{h}_{\mathrm{ora},\mathcal{R}^{c}}^{\star}\Vert_{2}^{2}\le2(1+\xi^{2})\gamma^{2}\cdot g(s)\cdot\mathbb{E}\Vert\mathbf{Ah}\Vert_{2}^{2}.
\end{equation}
Subtracting the lower bound from the upper bound of the MSE, i.e.,
subtracting \eqref{MSE_le_GSURE_Theta_G} from \eqref{MSE_ge_GSURE_Theta_MSE}
and using $\mathrm{GSURE}(\mathbf{h}_{\mathrm{G}}^{\star})\le\mathrm{GSURE}(\mathbf{h}_{\mathrm{ora}}^{\star})$,
we obtain
\begin{align}
\frac{\mathrm{MSE}_{\mathrm{G}}-\mathrm{MSE}_{\mathrm{ora}}}{\mathbb{E}\Vert\mathbf{Ah}\Vert_{2}^{2}}\le & \frac{N}{M}\frac{2\epsilon}{1-\epsilon^{2}}\frac{\mathrm{GSURE}(\mathbf{h}_{\mathrm{G}}^{\star})}{\mathbb{E}\Vert\mathbf{Ah}\Vert_{2}^{2}}\nonumber \\
 & +\frac{4(1+\xi^{2})\gamma^{2}\delta}{1-\epsilon^{2}}\frac{2N-T}{T}g(s).
\end{align}
Using
\begin{equation}
\mathrm{GSURE}(\mathbf{h}_{\mathrm{G}}^{\star})\le\beta^{2}\mathbb{E}\Vert\mathbf{Ah}\Vert_{2}^{2}\le\beta^{2}\frac{M}{N}(1+\delta)\mathbb{E}\Vert\mathbf{h}\Vert_{2}^{2}
\end{equation}
and letting
\begin{gather}
\epsilon_{1}=\frac{2\left(1+\frac{1}{\sqrt{\rho}}\right)}{1-\left(1+\frac{1}{\sqrt{\rho}}\right)^{2}\delta^{2}},\\
\epsilon_{2}=\frac{4\kappa\cdot(1-\rho)}{\rho}\frac{1+\xi^{2}}{1-\left(1+\frac{1}{\sqrt{\rho}}\right)^{2}\delta^{2}},
\end{gather}
with $\rho=\frac{T}{2N},\eta=\frac{M}{N}$, we can derive
\begin{equation}
\left|\mathrm{NMSE}_{\mathrm{G}}-\mathrm{NMSE}_{\mathrm{ora}}\right|\le\left(\epsilon_{1}\beta^{2}+\epsilon_{2}\gamma^{2}g(s)\right)\delta(1+\delta).
\end{equation}



\bibliographystyle{IEEEtran}
\bibliography{IEEEabrv,GSURE-DEQ}



\end{document}